\documentclass[reqno]{amsart}
\usepackage{amsmath,amssymb,cite}

\newtheorem{theorem}{Theorem}[section]   
\newtheorem{proposition}[theorem]{Proposition}
\newtheorem{lemma}[theorem]{Lemma}
\newtheorem{corollary}[theorem]{Corollary}

\numberwithin{equation}{section}
\numberwithin{theorem}{section}

\renewcommand{\(}{\left(}      \renewcommand{\)}{\right)}
      
   \newcommand{\nn}{\nonumber}  

\newcommand{\mc}[1]{{\mathcal #1}}

\newcommand{\bb}[1]{{\mathbb #1}}

\renewcommand{\epsilon}{\varepsilon}
\newcommand{\id}{{1 \mskip -5mu {\rm I}}}
 
\newcommand{\varsh}{\mathop{\rm sh}\nolimits}
\newcommand{\varch}{\mathop{\rm ch}\nolimits}
\newcommand{\varth}{\mathop{\rm th}\nolimits}
\newcommand{\sgn}{\mathop{\rm sgn}\nolimits}

\newcommand{\bam}{\,\overline {\!m\!}\,}
\newcommand{\la}{\lambda}

\begin{document}

\title[Soft and hard wall]
{Soft and hard wall in a stochastic \\ reaction diffusion equation}

\author [L.\ Bertini] {Lorenzo Bertini}
\address{Lorenzo Bertini, Dipartimento di Matematica,
Universit\`a di Roma `La Sapienza', P.le Aldo Moro 2,
00185 Roma, Italy}
\email{bertini@mat.uniroma1.it}

\author [S.\ Brassesco] {Stella Brassesco}
\address{Stella Brassesco, Departamento de Matem\'aticas, 
Instituto Venezolano de Investigaciones Cient\'{\i}ficas,
Apartado Postal 21827, Caracas 1020--A, Venezuela}
\email{sbrasses@ivic.ve}

\author [P.\ Butt\`a] {Paolo Butt\`a}
\address{Paolo Butt\`a, Dipartimento di Matematica,
Universit\`a di Roma `La Sapienza', P.le Aldo Moro 2,
00185 Roma, Italy}
\email{butta@mat.uniroma1.it}



\begin{abstract}
We consider a stochastically perturbed reaction diffusion equation in
a bounded interval, with boundary conditions imposing the two stable
phases at the endpoints.  We investigate the asymptotic behavior of 
the front separating the two stable phases, as the intensity
of the noise vanishes  and the size of the interval diverges. 
In particular, we prove that, in a suitable scaling limit,
the front evolves according to a one-dimensional diffusion process
with a non-linear drift accounting for a ``soft'' repulsion from the
boundary.  We finally show how a ``hard'' repulsion can be obtained by an extra diffusive scaling.
\end{abstract}

\maketitle

   \section{Introduction}
   \label{sec:1}
\thispagestyle{empty}

Let $V(m)$ be a smooth, symmetric, double well potential whose 
 minimum is 
attained at $m=m_\pm$, $V''( m_\pm)>0$. 
After the pioneering paper \cite{AC}, the semi-linear 
parabolic equation
   \begin{equation}
   \label{ac}
\frac{\partial m}{\partial t} =  
\frac12 \Delta m -V'(m) 
   \end{equation}
and its stochastic perturbations, have became a basic model in the
kinetics of phase separation and interface dynamics for systems with a
non conserved order parameter. 

Before introducing our results, let us review the main features
of \eqref{ac} in the one dimensional case. The 
corresponding evolution is the $L_2$ gradient flow of the functional
   \begin{equation}
   \label{acf}
\mathcal{F}(m) = \int\!dx\, \Big[ \frac 12 m'(x)^2
+ 2V(m(x)) \Big].
   \end{equation}
In the case that  \eqref{ac} is considered in the
whole line $\bb R$, there are infinitely many stationary solutions,
which are the critical points of $\mathcal{F}$. The most relevant
are the constant profiles $m_\pm$, where 
$\mathcal{F}$ attains its minimum, and $\pm\bam$, where $\bam$
is the  solution to  
   \begin{equation}
   \label{1.2.5}
\frac 12 \bam''- V'(\bam ) =0, \qquad \lim_{x\to
\pm\infty} \bam(x) = m_\pm, \qquad \bam(0) = 0,
   \end{equation}
together with its translates $\pm\bam_\zeta(x) =\pm \bam(x-\zeta)$,
$\zeta\in\bb R$. The profile $\bam_\zeta$ is a standing wave of
\eqref{ac} that connects the two pure phases $m_\pm$. Note that
$\bam_\zeta$ minimizes $\mathcal{F}$ under the constraint that
$\lim_{x\to \pm\infty} m(x) = m_\pm$. Therefore $\bam_\zeta$ is the
equilibrium state which has the two pure phases $m_\pm$ coexisting to
the right and to the left of $\zeta$. It represents a mesoscopic
interface located at $\zeta$. We use the word ``mesoscopic'' because
the interface is diffuse and the transition from one phase to the
other, even though exponentially fast, is not sharp. In \cite{FM} it
is proven that the one parameter invariant manifold $\mathcal{M} =
\{\bam_\zeta : \zeta\in {\bb R}\}$ is asymptotically stable for the
evolution \eqref{ac}. 

Referring \cite{Funaki} for a review on stochastic interface models,
we outline some results on the stochastic perturbation of \eqref{ac}.
When a random forcing term of intensity $\sqrt\epsilon$ is added to 
\eqref{ac} and the initial datum is $\bam_0$, in \cite{BDP,BBDP,funaki}
it is shown that the solution at times $\epsilon^{-1}t$ stays close 
to $\bam_{\zeta_\epsilon(t)}$ for some $\zeta_\epsilon(t)$ which
converges to a Brownian motion as $\epsilon\to 0$. To
explain heuristically this result, let us regard the random forcing
term as a source of independent small kicks, which we decompose along
the directions parallel and orthogonal to $\mathcal M$. The orthogonal
component is exponentially damped by the deterministic drift, while
the parallel component, associated to the zero eigenvalue of the
linearization of \eqref{ac} around $\bam_\zeta$, is not contrasted
and, by independence, sums up to a Brownian motion. 

We next discuss the behavior of Allen-Cahn equation on the bounded
interval $[-a,b]$. The case of Neumann boundary conditions is 
considered in \cite{CP,FH}, where it is shown that there
exists a stationary solution $m^*_{a,b}$, close to $\bam_{(b-a)/2}$ as
$a,b$ diverge. The profiles $\pm m^*_{a,b}$ are saddle points of
$\mathcal{F}$, each one having a one dimensional unstable manifold
connecting it to the stable points $m_\pm$. For $a,b$ large, solutions
are first attracted by this manifolds and they then move  along it 
toward one of the stable phases, with a velocity exponentially small 
in the distance from the endpoints. From the
analysis in \cite{CP,FH}, we have that there exists a constant $c_0>0$
(depending on the potential $V$) such that, if we take
$a=c_0\log\epsilon^{-1}$, $b = \epsilon^{-\beta}$ for some $\beta>0$,
and the initial condition is close to $\bam_0$, the following
holds. As $\epsilon\to 0$, the solution of \eqref{ac} at times
$\epsilon^{-1}t$, for $t$ small enough, is close to $\bam_{\zeta(t)}$,
where $\zeta(t)$ solves the equation $\dot \zeta = - A\, \epsilon^{-1}
e^{-(\zeta+a)/c_0} = - A\, e^{-\zeta/c_0}$ for some $A>0$. When a
random forcing term of order $\sqrt\epsilon$ is added to \eqref{ac},
by the analysis in \cite{BBDP}, it follows that, by taking
$a=c\log\epsilon^{-1}$ with $c\gg c_0$, and looking at the time scale
$\epsilon^{-1}$, the random fluctuations are dominant so that the limiting
motion of the interface is still described by a Brownian. On the other
hand, for $c < c_0$ the deterministic drift should become dominant,
the minority phase shrinking deterministically up to extinction. In
the critical case $c=c_0$, at the initial state of the evolution, we
should see the effect both of the drift and of the stochastic fluctuations.

In this paper, we consider a stochastic perturbation of \eqref{ac} in
a bounded interval with inhomogeneous Dirichlet boundary conditions
imposing the two stable phases $m_\pm$, and analyze the competition
between the stochastic fluctuations and the given boundary conditions on
the motion of the interface. Let us first consider the deterministic
case, that is, \eqref{ac} in the interval $[-a,b]$ with boundary
conditions $m(t,-a) = m_-$, $m(t,b)=m_+$. The meaning of these
conditions is to force the $m_-$ phase, respectively the $m_+$ phase,
to the left of $-a$, respectively to the right of $b$. If we think of $m$ as the local magnetization, this choice models the effect of opposite magnetic fields applied at the endpoints. To our knowledge, an
analysis along the same lines of \cite{CP,FH} has not been carried out
in detail. However, in this case, it is straightforward to check that
there exists a unique, globally attractive, stationary solution
$m^*_{a,b}$, close to $\bam_{(b-a)/2}$ as $a,b$ diverge. Moreover, as
it follows from the analysis of the present paper, there is a slow
motion as in the case of Neumann boundary conditions. More precisely,
there exists an approximately invariant manifold $\mathcal{M}_{a,b}$,
close to $\mathcal{M}$ as $a,b$ diverge. In this limit, the motion
near $\mathcal{M}_{a,b}$ can be described in terms of coordinates
along and transversal to $\mathcal{M}_{a,b}$. The transversal
component of the flow is exponentially damped uniformly in $a,b$,
while the motion along $\mathcal{M}_{a,b}$, parametrized by the
interface location $\zeta(t)$, evolves according to $\dot\zeta = A\,
\big[ e^{-(\zeta+a)/c_0} - e^{-(b-\zeta)/c_0} \big]$ for $A$ and $c_0$
positive constants.  We emphasize that, since the boundary conditions
force the presence of an interface, the drift pushes the solution
toward $m^*_{a,b}$, where the two pure phases coexist.

We consider a stochastic perturbation  of \eqref{ac},
 given by a space--time white noise of
intensity $\sqrt\epsilon$.  To get a nontrivial scaling limit, 
and to  see the competition between the random fluctuations and the repulsion 
from one endpoint ($-a$),   we choose
$a=c_0\log\epsilon^{-1}$, $b =\epsilon^{-\beta}$, for some $\beta>0$,
the initial condition close to $\bam_0$, and look at the evolution at
times $\epsilon^{-1}t$. We  prove that, as $\epsilon\to 0$, the
solution stays close to $\bam_{\zeta(t)}$, where $\zeta(t)$ solves the
stochastic equation $\dot \zeta = A\,e^{-\zeta/c_0} + \eta$, here
$\eta$ is a white noise. We interpret this result as a ``soft wall'',
since the repulsion is not sharp. Actually, the solution 
remains close to $\mathcal{M}_{a,b}$  also on a slightly longer time
scale and
performing a further diffusive rescaling of the interface location, we also
prove that the soft wall converges to a ``hard'' one: the interface
dynamics behaves as a reflected Brownian motion.

   \section{Notation and results}
   \label{sec:2}   

Let $a,b\in\bb R_+$,  $\big(\Omega, \mc F, {\mc F}_t, \bb P\big)$ 
be a standard filtered probability space, and $W=\{W(t), t\in\bb R_+\}$
be the cylindrical Wiener process on $L_2([-a,b],dx)$. This means that
$W$ is the $\mc F_t$-adapted mean zero Gaussian process such that, for
each $\varphi,\varphi'\in C^\infty([-a,b])$ and $t,t'\in\bb R_+$,
   \begin{equation}
   \label{cov}
\bb E \Big( 
\langle W(t),\varphi\rangle \langle W(t'),\varphi'\rangle \Big)
= t\wedge t' \: \langle\varphi,\varphi'\rangle,
   \end{equation}
where $\bb E$ denotes the expectation w.r.t.\ $\bb P$, 
$t\wedge t' := \min\{t,t'\}$, and $\langle \cdot,\cdot\rangle$ 
is the inner product in $L_2([-a,b], dx)$.

In this paper we consider the prototypical case of the symmetric
double well potential, i.e.\ we choose
    \begin{equation}
    \label{3.2a}
V(m) = \frac 14 \big(m^2-1\big)^2,
    \end{equation}
which attains its minimum at $m=\pm 1$. Given $\epsilon>0$, we consider 
a stochastic perturbation of the one dimensional reaction diffusion equation 
\eqref{ac} with inhomogeneous Dirichlet boundary conditions at the endpoints. 
More precisely, we let $m(t) \equiv m(t,x)$, $(t,x)\in\bb R_+\times
[-a,b]$ be the solution to  
  \begin{equation}
  \label{feq}
\begin{cases}
{\displaystyle 
d m(t)= \big[ \frac12\Delta m(t) -V'(m(t)) \big]dt  +\sqrt{\epsilon}\, dW(t),
}
\\
{\displaystyle 
m(t,-a)=-1,
}
\\
{\displaystyle 
m(t,b)=1,
}\\
{\displaystyle 
m(0,x)=m_0(x).
} 
\end{cases}
  \end{equation}
To give a precise meaning to the above equation for 
$m_0\in C([-a,b])$ such that $m_0(-a)=-1$ and $m_0(b)=1$, 
let $\nu(x)=\frac{2x}{a+b}+\frac{a-b}{a+b}$ be the 
solution of $\nu''(x)=0$, $x\in (-a,b)$ with the above boundary conditions
and denote by $p^0_{t}$ the heat semigroup on $(-a,b)$ with zero
boundary conditions at the endpoints. Then a mild solution 
of  \eqref{feq} is defined as 
the solution of the integral equation
 \begin{equation}
\label{ieq}
m(t)=\nu + p^0_{t}(m_0-\nu)
-\int_0^t\, ds\, p^0_{t-s}V'(m(s)) +\sqrt{\epsilon} \int^t_0 p^0_{t-s}dW(s).
\end{equation}
By e.g.\ \cite{JF}, there exists a unique $\mc F_t$-adapted process
$m\in C(\bb R_+;C([-a,b])$ which solves \eqref{ieq}.

As explained in the Introduction, let $\bam_\zeta(x)$ be the 
standing wave with ``center'' $\zeta\in \bb R$, i.e.\ the
solution to \eqref{1.2.5}. For the specific choice potential 
\eqref{3.2a} of the potential we have $\bam_\zeta (x) =
\varth (x-\zeta)$.
Note that, if $a=b=\infty$ and $\epsilon=0$, then $\mc M = \{\bam_\zeta,\zeta\in\bb R\}$ is a one parameter
family of stationary solutions of \eqref{feq}.  Given $p\in
[1,\infty]$ we denote by $\|\cdot\|_p$ the norm in
$L_p([-a,b],dx)$. We consider $C(\bb R_+)$ equipped with the
(metrizable) topology of uniform convergence in compacts.  Our main
results are stated as follows.

\begin{theorem}
  \label{t:1}
Given $\beta >0$, set 
   \begin{equation}
\label{eq:1.1}
a := \frac 14 \log \epsilon^{-1},
\qquad b := \epsilon^{-\beta},
\qquad \lambda := \log\epsilon^{-1},
   \end{equation}
and denote by $m^{(\epsilon)}(t)$ the solution to \eqref{ieq} with
initial datum $m_0^{(\epsilon)} \in C([-a,b])$,
$m_0^{(\epsilon)}(-a) = -1$, $m_0^{(\epsilon)}(b) =1$, such that for
each $\eta>0$ we have
\begin{equation}
  \label{eq:m0}
  \lim_{\epsilon\to 0} 
  \epsilon^{-\frac 12 +\eta} \big\|m_0^{(\epsilon)}-\bam_0\big\|_\infty  = 0.
\end{equation}
Then:
\begin{enumerate}
\item[(\textit{i})]{
there exists a $\mc F_t$-adapted real process $X_\epsilon$ such
that, for each $\theta,\eta>0$,
\begin{equation}
  \label{eq:1.2}
\lim_{\epsilon\to 0} \bb P \Big( 
\sup_{t\in [0,\lambda \epsilon^{-1}\theta]} \big\|
m^{(\epsilon)}(t)-\bam_{X_\epsilon(t)}\big\|_{\infty} > 
\epsilon^{\frac 12 -\eta} \Big) = 0;
\end{equation}
}
\item[(\textit{ii})]{
the real process $Y_\epsilon(\tau):= X_\epsilon(\epsilon^{-1}\tau)$, $\tau
\in \bb R_+$, converges  weakly in $C(\bb R_+)$ to the unique strong solution
$Y$ to the stochastic equation
\begin{equation}
  \label{eq:1.3}
  \begin{cases}
    {\displaystyle   
      dY(\tau)= 12 \exp\{ - 4 Y(\tau) \}  d\tau + d B(\tau),
      }
      \\
      {\displaystyle   
      Y(0) = 0,
      }
   \end{cases}  
\end{equation}
where $B$ is a Brownian motion with diffusion coefficient $\frac 34$;
}
\item[(\textit{iii})]{
the real process $Z_\epsilon(\theta):= \lambda^{-1/2}
X_\epsilon(\lambda\epsilon^{-1}\theta)$, $\theta\in \bb R_+$, converges
weakly in $C(\bb R_+)$ to a Brownian motion with diffusion coefficient
$\frac 34$ reflected at zero.}
\end{enumerate}
\end{theorem}

Item (\textit{i})  states
that, up to times $\epsilon^{-1}\log\epsilon^{-1}$, the
solution of \eqref{feq} with initial condition close to 
the one-dimensional manifold $\{\bam_\zeta;\zeta\in (-a,b)\}$
remains close to that manifold. 
Items (\textit{ii}) and (\textit{iii}) then identify the limiting
evolution of the interface $X_\epsilon(t)$. On the time scales
$\epsilon^{-1}$ the interface is at distance $\frac 14 \log\epsilon^{-1} +
Y_\epsilon$ from the endpoint $-a$; moreover $Y_\epsilon$ behaves as a
Brownian motion with a strong drift toward the right for
$Y_\epsilon<0$ and essentially no drift for $Y_\epsilon>0$. We
interpret this as a ``soft wall''. On the longer time scale
$\epsilon^{-1}\log\epsilon^{-1}$ the interface is at distance
$\frac 14 \log\epsilon^{-1} + \sqrt{\log\epsilon^{-1}} Z_\epsilon$
from the endpoint $-a$; on this time scale the repulsion is sharp:
$Z_\epsilon$ behaves as a  Brownian motion reflected at zero.  We
interpret this as a ``hard wall''. We finally remark that the choice
of $\lambda$ in \eqref{eq:1.1} has been made for the sake of
concreteness: it would have been enough to take $\lambda$
such that $\lambda\to \infty$ and $\sqrt\lambda/\log\epsilon^{-1}
\to 0$ as $\epsilon\to 0$.

We emphasize that this nontrivial behavior is due to the choice
$a=\frac 14 \log\epsilon^{-1}$ for which there is a competition
between the stochastic fluctuations and the drift due to the Dirichlet
boundary condition at the endpoint $-a$. Here the coefficient $\frac
14$, as well as the diffusion coefficient $\frac 34$ of the Brownian
motion, depend on the special choice of the double well potential $V$
in \eqref{3.2a}. Since $b=\epsilon^{-\beta}\gg a$ the right endpoint
$b$ has not effect on the limiting motion of the interface; apart from 
(minor) technical details the case $b = +\infty$ behaves as the one
here considered. It follows from our analysis that if we had chosen
 $a=(\frac 14 +\delta)
\log\epsilon^{-1}$
  for some $\delta>0$, in the limiting motion of the
interface we would have seen only the effect of the stochastic force,
namely $Y_\epsilon$ would behave as a Brownian motion. On the other
hand, if we had chosen $a=(\frac 14 -\delta)\log\epsilon^{-1}$ we
would have felt, by looking at the initial stage of the evolution on
the time scale $\epsilon^{-1}$, an infinite drift toward the right. In
such a situation it should be possible to show that the process 
$Y_\epsilon(t)$, $t\in (0,\infty)$, still converges to a solution to 
the stochastic equation in \eqref{eq:1.3}.

\medskip
In \cite{bbbeq} we analyze the invariant measure $\mu_\epsilon$
of \eqref{feq} with $a=b=\frac 14 \log\epsilon^{-1}$ and show it has 
a nontrivial limit as $\epsilon\to 0$. In fact in 
\cite{bbbeq} the main effort is in proving the compactness of 
$\mu_\epsilon$, relying in the following dynamical scaling
limit to identify its limit points. Fix $\tau_0$ and let
$\bb Q^\epsilon_{m_0}$ be the law of $m(\epsilon^{-1}\tau)$, 
$\tau\in [0,\tau_0]$, with $m(t)$ the
solution to \eqref{feq} with $a=b=\frac 14 \log\epsilon^{-1}$. 
By setting $m(t,x) = \sgn(x)$ for $|x|\ge \frac 14 \log\epsilon^{-1}$,
we regard $\bb Q^\epsilon_{m_0}$ as a probability measure on 
$C([0,\tau_0]; \mc X)$ where $\mc X := 
\big\{m\in C(\bb R) \, : \, \lim_{x\to\pm\infty} m(x) = \pm 1 \big\}$
endowed with the topology of uniform convergence.
Given $z\in\bb R$, we also let $\bb Q_z$ be the probability measure
on $C([0,\tau_0];\mc X)$ defined by $\bb Q_z(A) := 
P\big(\bam_{\Xi^z(\cdot)} \in A \big)$, where $\Xi^z(\tau)$, $\tau\in
[0,\tau_0]$ is the unique strong solution to the stochastic 
differential equation
   \begin{equation}
   \label{eqsim}
\begin{cases}
d\,\Xi(\tau) = - \, 24\, \varsh\big(4\,\Xi(\tau)\big) d\tau + d B(\tau), 
\\ \Xi(0)=z, \end{cases}
   \end{equation}
where $B$ is a Brownian motion with diffusion coefficient $\frac 34$.
Note that, although the drift term is not globally Lipschitz, a standard
coercivity argument shows the existence and uniqueness of 
the strong solution to \eqref{eqsim}. 
In this setting, the analogous of the convergence to the soft wall 
in Theorem~\ref{t:1} is the weak convergence of $\bb Q^\epsilon_{m_0}$
to $\bb Q_{z_0}$; here $m_0$ satisfies $\|m_0-\bam_{z_0}\|_\infty \le
\epsilon^{\frac 12 - \eta}$ for some $\eta$ small enough. In
\cite{bbbeq} we also need such convergence to hold uniformly for $z_0$
in compacts; this is the content of the following theorem.

\begin{theorem}
\label{t:10}
Let $\tau_0>0$. There exists $\eta_1>0$ such that for any 
$\eta\in [0,\eta_1]$ the following  holds. For each $L>0$ and each
uniformly continuos and bounded function 
$F : C([0,\tau_0];\mc X) \to \bb R$ we have
\begin{equation}
  \label{eq:1.31}
\lim_{\epsilon\to 0} \:\: \sup_{z\in [-L,L]} \:\:
\sup_{m_0\in \mc N^\epsilon_\eta(z)} \:\: \big| \bb Q^\epsilon_{m_0}
(F) - \bb Q_{z}(F) \big| = 0,
\end{equation}
where $\mc N^\epsilon_\eta(z) := \big\{m \in \mc X_\epsilon \, :\,
\|m-\bam_{z}\|_\infty \le \epsilon^{\frac 12 - \eta} \big\}$.
\end{theorem}

\medskip
\noindent\textit{Outline and basic strategy.} The proof of
Theorem~\ref{t:1} relies on an iterative scheme, in which we linearize
\eqref{ieq} around $\bam_\zeta$ for a suitable $\zeta$ recursively
defined. From a geometrical point of view, we approximate the flow
induced by \eqref{ieq} with a piecewise linear one, which stays close
to the quasi-invariant manifold $\mathcal{M}_{a,b}$, and allows to
compute the motion along the manifold itself. More precisely,
following \cite{BBBP1,BBDP,BDP}, we split the time axis into intervals
of length $T$, taking $T$ diverging as $\epsilon\to 0$, yet very small
as compared to the macroscopic time $\epsilon^{-1}$. For the piecewise
linear flow, we compute the displacement of the center, effectively
tracking the motion along the quasi-invariant manifold. To this end,
sharp estimates on the linear flow are needed. We emphasize that, even
if the linearization of \eqref{ac} on the whole line around the
standing wave $\bam_\zeta$ is very well understood \cite{FM}, for our
purposes the finite size corrections are crucial, the nonlinear drift
in \eqref{eq:1.3} being indeed due to them. Moreover, to control the
difference between the true flow and the piecewise linear one, we need
\textit{a priori} bounds which allow us to neglect the nonlinear
terms. Finally, the convergence to the hard wall stated in item
(\textit{iii}) is proven by showing that the interface motion is
accurately described by \eqref{eq:1.3} also on the time scale
$\lambda\epsilon^{-1}$. The proof then follows by showing that the
diffusive scaling of the latter converges to a reflected Brownian motion. The proof of Theorem~\ref{t:10} requires only minor modifications and it is sketched in Appendix~\ref{s:A}. 

   \section{The iterative scheme} \label{sec:3}

The notion of  ``center'' of a function plays an important role in
our analysis. Following \cite{BDP,BBDP}, given a function $f\in
C([-a,b])$ we define its center $\zeta$ as a point in $(-a,b)$ such
that
\begin{equation}
   \label{defcent}
\langle f - \bam_{\zeta}\,,\,\bam'_{\zeta}\rangle =
\int^b_{-a}\!dx\, \big[ f(x)- \bam_\zeta(x) \big] \,\bam '_\zeta(x) = 0.
  \end{equation}
Referring to \cite{BDP} for an interpretation of the above
definition in terms of the dynamics given by the linearization of
\eqref{feq} around $\bam_{\zeta}$, here we simply note that $\zeta$ 
minimizes the $L_2$ norm of $f-\bam_{z}$ as a function of $z$.

Given  $\delta,\ell>0$ we define
   \begin{equation}
\label{defcset}
\Upsilon (\delta,\ell)
:= \Big\{ f \in C([-a,b])\, :\: 
\big\| f -\bam_{z} \big\|_\infty < \delta 
\mbox{ for some } z \in (-a+ \ell ,b-\ell) 
\Big\}.
   \end{equation}

Existence and uniqueness of the center holds for functions in
$\Upsilon (\delta,\ell)$
 for $\epsilon,\delta$ small enough and
$\ell$ large enough, as precisely stated in the next proposition.
Recall that we have chosen $a=\frac 14 \log\epsilon^{-1}$, $b =
\epsilon^{-\beta}$. The result is analogous to \cite[Prop.~3.2]{BBDP} where the whole line is considered, and the proof follows
by standard implicit function arguments \cite{BDP,BBDP}.
 
\begin{proposition}
\label{prop2.1}
There are reals $\delta_0,\ell_0>0$ such that, for any 
$\epsilon$ small enough, if  $f\in \Upsilon (\delta_0,\ell_0)$
then $f$ has a unique center $\zeta\in (-a,b)$.  
Moreover there is a constant $C_0>0$ so that 
if $z\in (-a+\ell_0,b-\ell_0)$ is such that 
$\big\| f -\bam_{z} \big\|_\infty <\delta_0$ we have
   \begin{equation*}
|\zeta-z|\le C_0 \big\| f -\bam_{z} \big\|_\infty
\end{equation*}
and 
   \begin{eqnarray*}
&& \zeta = z - \frac 34 \big\langle\bam '_{z},f - \bam_z\big\rangle
- \frac{9}{16} \big\langle\bam'_{z},f - \bam_{z}\big\rangle
\big\langle\bam''_{z},f - \bam_{z}\big\rangle
+ R(z,f),
\\ && 
| R(z,f) | \le 
C_0 \Big\{ \big\| f -\bam_{z} \big\|_\infty^3 
+ \big(e^{-2(b-z)}+e^{-2(a+z)}\big)\, \big\| f -\bam_{z} \big\|_\infty  \Big\}.
   \end{eqnarray*}
   \end{proposition}

In the sequel, given $f\in\Upsilon (\delta,\ell)$ with $\delta<\delta_0$ and 
$\ell > \ell_0$, we denote by $X(f)$ the center of $f$, which is well defined
for $\epsilon$ sufficiently small. From now on we drop however
the explicit dependence on $\epsilon$ from the notation. Let $m(t)$ be the 
solution to \eqref{ieq} with $m_0$ satisfying \eqref{eq:m0} and $\alpha\in 
(0,1)$; we define the stopping times
\begin{eqnarray}
  \label{eq:sdl}
  S_{\delta,\ell}
  &:=&\inf\big\{t\in\bb R_+ \,:\, m(t)\not\in \Upsilon (\delta,\ell) \big\},
  \\
  \label{eq:sdla}
  S_{\delta,\ell,\alpha}
  &:=& S_{\delta,\ell} \land \inf\big\{ t \, :\, 
  \big|X(m(t))\big| \ge \alpha \, a \big\}.
\end{eqnarray}

We analyze $m(t)$ as long as it stays in $\Upsilon(\delta,\ell)$ and
its center is not too far from the origin, namely we stop the
evolution at the time $S_{\delta,\ell,\alpha}$ by considering
$m(t\wedge S_{\delta,\ell,\alpha})$. We are going to  introduce an iterative
procedure in which we linearize the equation \eqref{ieq} around
$\bam_x$ for a suitable $x$ recursively defined. To do so, we need 
 a few definitions. 
 
Given $\zeta\in (-a,b)$, let $\varphi_\zeta \in C^2([-a,b])$ be the solution to
   \begin{equation}
   \label {eqphi}
\begin{cases}
{\displaystyle 
\frac 12 \varphi_\zeta''(x) - V''(\bam_\zeta(x)) 
\varphi_\zeta(x)  = 0, } 
\\
{\displaystyle \varphi_\zeta(-a)=-1-\bam_{\zeta}(-a),
}\\
{\displaystyle \varphi_\zeta(b) = 1-\bam_{\zeta}(b). 
}
   \end{cases}
   \end{equation}
An explicit computation yields 
   \begin{equation}
   \label {exphi}
\varphi_\zeta (x)=\bam'_{\zeta}(x) \big[c_\zeta  q_\zeta(x)  +d_\zeta\big],
\qquad
q_\zeta(x) := \frac{h_\zeta(x) - h_\zeta(-a)}{h_\zeta(b) - h_\zeta(-a)},
   \end{equation} 
where
   \begin{eqnarray}
   \label{eq:2.1}
&& 
h_\zeta(x) := \int_{\zeta}^x\!dy\, \frac{1}{\bam'_{\zeta}(y)^2}
 \,=\,\frac 38 (x-\zeta)\,+\,\frac 38 
\frac{\bam_\zeta(x)}{\bam_\zeta'(x)}\,+\,
\frac14 \frac{\bam_\zeta(x)}{\bam_\zeta'(x)^2},
\phantom{amor}
\\
   \label{eq:2.1bis}
&&
c_\zeta :=  \frac{1}{1-\bam_\zeta(-a)} + \frac{1}{1+\bam_\zeta (b)}, 
\qquad d_\zeta := -\frac{1}{1-\bam_\zeta(-a)}.
   \end{eqnarray}

We also introduce the operator $H_\zeta$ on $C_0([-a,b])$, the space of
continuous functions vanishing at the endpoints, defined on $C^2_K([-a,b])$,
the space of twice differentiable functions compactly supported in $(-a,b)$,
by   
\begin{equation}
   \label{dhz}
   H_\zeta f (x) := - \frac 12 f''(x) + V''(\bam_\zeta(x)) f(x)
\end{equation}
and denote by $g_t^{(\zeta)} := \exp\{-tH_\zeta\}$ the corresponding
semigroup.

Let $t_0\in\bb R_+$, and $m(t)$, $t\ge t_0$ be the solution to
\eqref{feq} with initial condition $m(t_0)=\bam_{\zeta}+ \vartheta$,
for some $\zeta\in (-a+\ell_0,b-\ell_0)$ and $\vartheta\in C([-a,b])$ such
that $\|\vartheta\|_\infty < \delta_0$ 
($\delta_0,\ell_0$ as in Proposition~\ref{prop2.1}). By
writing $m(t)=\bam_\zeta + v(t)$ and expanding $V'(\bam_\zeta+v)=
V'(\bam_\zeta)+ V''(\bam_\zeta)
v + 3 \bam_\zeta v^2 + v^3$, it is
easy to check from \eqref{feq} that $v(t)$ satisfies the integral equation
\begin{equation}
\label{iequ}
v(t)= \varphi_\zeta 
+  g_{t-t_0}^{(\zeta)} \big( \vartheta -\varphi_\zeta\big)
-\int_{t_0}^t\!ds\, g_{t-s}^{(\zeta)}
\big[ 3\bam_\zeta v(s)^2 + v(s)^3 \big]
+\sqrt{\epsilon} \int_{t_0}^t g_{t-s}^{(\zeta)} dW(s).
\end{equation}

Let now $m(t)$, $t\ge 0$, be the solution to \eqref{feq} and  
consider the partition $\bb R_+=\bigcup _{n\ge0}[T_n,T_{n+1})$,
 where $T_n
= nT$, $n\in \bb N$ and $T= \epsilon^{-\gamma}$, $\gamma\in 
\big(0,\frac 18\big)$.
We next define, by induction on $n\ge 0$, reals $x_n$ and functions
$v_n(t) \equiv \{v_n(t,x), \, x\in [-a,b]\}$, $t \in
[T_n,T_{n+1}]$. They will have the property that for any
$t\in[T_n,T_{n+1}]$
   \begin{equation}
   \label{4.1}
m\big( t\wedge S_{\delta,\ell,\alpha} \big)
= \bam_{x_n} + v_n (t) 
   \end{equation}

Set $x_0:=X(m_0)$, i.e.\ the center of $m_0$, and let $v_0(t)$,
$t\in[0,T]$, be the solution to \eqref{iequ} with $t_0=0$,
$\zeta=x_0$, and $\vartheta=m_0 - \bam_{x_0}$, stopped at
$S_{\delta,\ell,\alpha}$.  Suppose now, by induction, that we have
defined $x_{n-1}$ and $v_{n-1}$. We then define $x_n$ as the
center of $m(T_n \wedge S_{\delta,\ell,\alpha})=\bam_{x_{n-1}}+
v_{n-1}(T_n)$ (which exists by the definition of the stopping time
$S_{\delta,\ell,\alpha}$) and $v_n(t)$, $t\in [T_n,T_{n+1}]$, as the
solution to \eqref{iequ} with $t_0=T_n$, $\zeta=x_n$, and $\vartheta=
m(T_n \wedge S_{\delta,\ell,\alpha})-\bam_{x_n}$, stopped at
$S_{\delta,\ell,\alpha}$.  We emphasize that in this construction the
initial condition $v_n(T_n)$ for the evolution in the interval
$[T_n,T_{n+1}]$ is related to the final condition $v_{n-1}(T_n)$ of
the previous interval by
\begin{equation}
  \label{4.2}
v_{n}(T_{n})=-\bam_{x_{n}} + \bam_{x_{n-1}}+ v_{n-1}(T_{n}) 
\end{equation}  
  
We consider the operator $H_\zeta$ defined in \eqref{dhz} also as an
operator on $L_2([-a,b],dx)$ self-adjoint with domain
$W^{2,2}([-a,b],dx)\cap W^{1,2}_0([-a,b],dx) $. The bottom of its
spectrum is an isolated eigenvalue $\lambda_0^{(\zeta)}>0$ of
multiplicity one. The corresponding eigenfunction, that we denote by
$\Psi_0^{(\zeta)}$, is chosen positive. We also introduce the
\emph{spectral gap} of $H_\zeta$ which is defined as
$\mathrm{gap}(H_\zeta) := \inf \mathrm{spec}(H_\zeta\restriction
(\Psi_0^{(\zeta)})^\perp)$, where $H_\zeta\restriction
(\Psi_0^{(\zeta)})^\perp$ denotes the restriction of $H_\zeta$ to the
subspace orthogonal to $\Psi_0^{(\zeta)}$. Recalling $g^{(\zeta)}_t = 
e^{-tH_\zeta}$, we then define
   \begin{eqnarray}
   \label{gperp}   
g^{(\zeta,\perp)}_t f &:= & 
g^{(\zeta)}_t f -  e^{-\lambda_0^{(\zeta)} t} 
\big\langle \Psi_0^{(\zeta)}, f \big\rangle \,  \Psi_0^{(\zeta)},
\\
\label{Gperp}
G^{(\zeta,\perp)} &:= & 
\int_0^\infty\!dt \, g^{(\zeta,\perp)}_t.
   \end{eqnarray}
Note that $G^{(\zeta,\perp)}$ is well defined as
$\lambda_0^{(\zeta)}>0$. We denote by $g^{(\zeta,\perp)}_t(x,y)$, $t>0$, 
and $G^{(\zeta,\perp)}(x,y)$, $x,y\in[-a,b]$, the corresponding integral 
kernels. We shall use the same notation for the semigroups acting 
on $C([-a,b])$.

Let $\,\overline {\! H\!}_\zeta$ be the same operator as in \eqref{dhz},
but defined on the whole line $\bb R$, i.e.\ as an operator on $C_{\rm
b}(\bb R)$, the space of bounded continuous functions, or on $L_2(\bb
R, dx)$. It is well known that $\,\overline {\! H\!}_\zeta$ has a zero
eigenvalue with eigenfunction $\bam_\zeta'$ and a strictly positive
spectral gap \cite{FM}. This properties play a crucial role in the
analysis of the interface fluctuations for a stochastic reaction
diffusion equation on the whole line or, in any case, with the
interface sufficiently far from the boundary, see
\cite{BBBP1,PS,BBDP,BDP,funaki}. Analogously, we need sharp bounds on
the convergence, in a suitable sense, of $H_\zeta$ to $\,\overline {\!
H\!}_\zeta$ as $\epsilon \to 0$, which are stated below and proved in
Section \ref{sec:9}. Note that, since $H_\zeta$ and $\,\overline {\!
H\!}_\zeta$ are defined in different spaces, these bounds do not
follow directly from standard perturbation theory. We introduce
   \begin{equation}
   \label{qa} 
\phi^{(\zeta)}(x) :=  \frac{\bam_{\zeta}'(x)}{\|\bam_\zeta'\|_2}, \qquad
x\in [-a,b].
   \end{equation} 

   \begin{theorem}
   \label{specst}
Set $a$ and $b$ as in the statement 
of Theorem~\ref{t:1}. Then, for each $\alpha\in (0,1)$ there exist 
reals $\epsilon_1, \delta_1,C_1>0$ 
such that, for any $\epsilon\in (0,\epsilon_1]$, $|\zeta| < \alpha a$, 
and $f\in C([-a,b])$
   \begin{eqnarray}
   \label{gin}
{\displaystyle 
\| g_t^{(\zeta)} f \|_\infty }
&\le & {\displaystyle C_1 \| f \|_\infty \quad for\,any \quad t\ge 0},
\\ \label{gap}
\mathrm{gap}(H_\zeta) & \ge & \delta_1,
\\ \label{gperpin}
{\displaystyle  \| g_t^{(\zeta,\perp)} f \|_\infty} 
&\le & {\displaystyle  C_1 \, e^{-\delta_1 t} \, \| f \|_2^{2/3} 
\|f\|_\infty^{1/3}\quad for\,any \quad t\ge 1}.
   \end{eqnarray}
Moreover, for each $\eta>0$,
   \begin{eqnarray}
   \label{psi0}
&& \lim_{\epsilon \to 0} \, \sup_{|\zeta| < \alpha a}
\epsilon^{-\frac32 (1-\alpha)+\eta}  \:
\big |\lambda_0^{(\zeta)} - 24\, \epsilon \, e^{-4\zeta} \big| = 0, 
\\ \label{psi2} && \lim_{\epsilon \to 0} \, 
\sup_{|\zeta| < \alpha a} 
\epsilon^{-{\frac{1-\alpha}2}+\eta}
\: 
\big\| \Psi_0^{(\zeta)} - \phi^{(\zeta)} \big\|_\infty = 0, 
\\ \label{psi1} && \lim_{\epsilon \to 0} \, 
\sup_{|\zeta| < \alpha a} 
\epsilon^{-{\frac{1-\alpha}2}+\eta}  
\:\big\| \Psi_0^{(\zeta)} - \phi^{(\zeta)} \big\|_1 = 0,
\\ 
\label{psi3} && 
\lim_{\epsilon \to 0} \, 
\sup_{|\zeta| < \alpha a} 
\epsilon^{-(1-\alpha)+\eta}  \:
\big\langle \big| \Psi_0^{(\zeta)} - \phi^{(\zeta)} \big|, 
\phi^{(\zeta)} \big\rangle = 0,
\\
\label{psi4b} &&
\lim_{\epsilon \to 0} \, \sup_{|\zeta| < \alpha a} 
\epsilon^{-\frac 12 (1-\alpha)+\eta}  
\Big| \int_{-a}^b\!dx\,\bam_\zeta'(x)\, \bam_\zeta(x)\,
G^{(\zeta,\perp)}(x,x)\Big| = 0.
   \end{eqnarray}
   \end{theorem}

   \section{A priori bounds}
   \label{sec:4}

The following lemma captures the correct asymptotic behavior of the
first terms on the r.h.s.\ of \eqref{iequ}. Recall that 
$T=\epsilon^{-\gamma}$, $\gamma\in \big(0,\frac 18\big)$ and that 
$\varphi_\zeta$ is defined in  \eqref{exphi}. 

   \begin{lemma}
   \label{ape1}
Let $\alpha\in (0,\gamma)$. Then for each $\eta>0$ 
   \begin{equation}
   \label{stimphi}
\lim_{\epsilon\to 0}\, \sup_{|\zeta| < \alpha a} \;
\sup_{t\in [0,T]} \, \epsilon^{-\frac 12 (1-\alpha) +\eta} \,
\big\|\varphi_\zeta - g_t^{(\zeta)}\varphi_\zeta \big\|_\infty = 0.
   \end{equation}
   \end{lemma}

\proof Recalling \eqref{exphi}--\eqref{eq:2.1bis} and \eqref{qa}  we write
    $$
\varphi_\zeta - g_t^{(\zeta)}\varphi_\zeta = c_\zeta \big[
\bam'_{\zeta}q_\zeta - g_t^{(\zeta)}(\bam'_{\zeta}q_\zeta) \big]
+ d_\zeta \, \big\|\bam_\zeta'\big\|_2 \,
\big[\phi^{(\zeta)} - g_t^{(\zeta)}\phi^{(\zeta)} \big].
    $$
Note that  $\|\bam_\zeta'\|_2^2 \le
\int_{-\infty}^\infty\!dx\,\bam_0'(x)^2 = \frac 43$ and
$\max\{|c_\zeta|;|d_\zeta|\} \le 1$ for $\epsilon$ sufficiently small, 
so from  \eqref{gin} we get
   \begin{equation*}
\big\|\varphi_\zeta - g_t^{(\zeta)}\varphi_\zeta \big\|_\infty
\le (1+C_1)\, \big\|\bam'_\zeta\, q_\zeta\big\|_\infty + 
\sqrt{\frac 43}\, 
\big\| \phi^{(\zeta)}- g_t^{(\zeta)}\phi^{(\zeta)} \big\|_\infty.
   \end{equation*}
By using $g^{(\zeta)}_t \Psi_0^{(\zeta)} = e^{ -\lambda_0^{(\zeta)} t}
\Psi_0^{(\zeta)}$ and again \eqref{gin},
   $$
\big\| \phi^{(\zeta)}- g_t^{(\zeta)}\phi^{(\zeta)} \big\|_\infty 
\le (1+C_1) \big\| \Psi_0^{(\zeta)} - \phi^{(\zeta)} \big\|_\infty +
\big(1-e^{ -\lambda_0^{(\zeta)} t}\big) \, 
\big\| \Psi_0^{(\zeta)} \big\|_\infty.
   $$
By \eqref{psi0}, for each $\eta>0$, we have $1-e^{ -\lambda_0^{(\zeta)} t} 
\le \epsilon^{1-\alpha-\eta}T$ for any $t\in [0,T]$, $|\zeta| < \alpha a$,
and $\epsilon$ small enough. Then, using \eqref{psi2},
   \begin{equation}
        \label{vt11}
\lim_{\epsilon\to 0}\, \sup_{|\zeta| < \alpha a} 
\sup_{t\in [0,T]} \, \epsilon^{-\frac 12 (1-\alpha) + \eta} \,
\big\| \phi^{(\zeta)}- g_t^{(\zeta)}\phi^{(\zeta)} \big\|_\infty = 0.
   \end{equation}
We next note that, by \eqref{eq:2.1}, there exists $C_2>0$ such that 
   \begin{equation}
   \label{ch}
\sup_{|\zeta| < \alpha a} \sup_{x\in [-a,b]}\;\;
\bam_\zeta'(x)^2 |h_\zeta(x)| \le C_2,
   \end{equation}
whence there is $C_3>0$ such that, for $|\zeta| < \alpha a$ and 
$\epsilon$ small enough,
   \begin{equation}
   \label{vt1}
\big|\bam'_\zeta(x) q_\zeta(x) \big| \le C_3
\frac{\bam_\zeta'(x)^{-1} - h_\zeta(-a) }{h_\zeta(b) - h_\zeta(-a)}
\le C \epsilon^{-\frac\alpha 2} \exp\big\{-2\,\epsilon^{-\beta}\big\},
   \end{equation}
where we used that
for $\epsilon$ small enough and  
$|\zeta| < \alpha a$,  $\bam_\zeta'(x)^{-1}$
  achieves its maximum at $x = b$. The estimate 
\eqref{stimphi} follows.
\qed

\medskip
To simplify the notation let us introduce, for $n\in\bb N$ and 
$t\in [T_n,T_{n+1}]$, 
   \begin{equation} 
   \label{defz}
z_n(t) := \int_{T_n}^t\! g_{t-s}^{(x_n)}\,dW(s),
   \end{equation} 
which is the last term that appears in the integral equation for $v_n$,
see \eqref{iequ}.   
Given $\tau\in\bb R_+$ we let $n_{\epsilon}(\tau):=[\epsilon^{-1}\tau/T]$
and $n_{\epsilon, \delta,\ell,\alpha}(\tau):=
\big[\big(\epsilon^{-1}\tau\wedge S_{\delta,\ell,\alpha}\big)/T\big]$. 
Given $\eta>0$, $\theta\in\bb R_+$, we define the event
   \begin{equation} 
   \label{Bdef}
\mathcal{B}_{\epsilon,\theta,\eta}^{(1)} := 
\Big\{ 
\sup_{0\le n \le n_\epsilon(\lambda\theta)} \; 
\sup_{t\in [T_n,T_{n+1}]} \| z_n(t) \|_\infty 
\le \epsilon^{-\eta} \sqrt T \Big\}.
   \end{equation}
Let also
   \begin{eqnarray}
   \label{zmaltese}
z_n^\perp (t) & := & z_n(t) - 
\big\langle \Psi_0^{(x_n)}, z_n(t) \big\rangle 
\, \Psi_0^{(x_n)} = \int_{T_n}^t\! g_{t-s}^{(x_n,\perp)}\, dW(s),
\\ \label{vmaltese}
v_n^\perp(t) &:= & v_n(t) - \big\langle \Psi_0^{(x_n)},v_n(t) \big\rangle\,
\Psi_0^{(x_n)}
    \end{eqnarray}
be the component of $z_n(t)$, resp.\ $v_n(t)$, orthogonal to 
$\Psi_0^{(x_n)}$. We define 
   \begin{equation} 
   \label{Bdef3}
\mathcal{B}_{\epsilon,\theta,\eta}^{(2)} := 
\left\{\sup_{0\le n\le n_\epsilon(\lambda\theta)} \;
\sup_{t\in[T_{n},T_{n+1}]} 
\|z_n^\perp(t)\|_\infty \le \epsilon^{-\eta} \right\}
   \end{equation}
and set $\mathcal{B}_{\epsilon,\theta,\eta} :=
\mathcal{B}_{\epsilon,\theta,\eta}^{(1)} \cap
\mathcal{B}_{\epsilon,\theta,\eta}^{(2)}$.  By standard Gaussian
estimates, see \cite[Appendix B]{BBBP1}, we have that for each
$\theta,\eta,q>0$
    \begin{equation}
       \label{gestperp}
\bb P\big(\mathcal{B}_{\epsilon,\theta,\eta} \big) \ge 1 - \epsilon ^q
     \end{equation}
for any $\epsilon$ small enough. 

   \begin{theorem}
   \label{apestimate}
Let $\alpha\in (0,\gamma)$; then there exists $\eta_0>0$ such that,
for any $\theta\in \bb R_+$ and  $\eta \in (0,\eta_0)$,  on the event 
$\mathcal{B}_{\epsilon,\theta,\eta}$ we have
   \begin{align}
   \label{ape}
& 
\sup_{0\le n \le n_\epsilon(\lambda\theta)} \; 
\sup_{t\in [T_n,T_{n+1}]} \| v_n(t) \|_\infty
\le \sqrt{\epsilon T} \, \epsilon^{-2\eta},
\\
\label{1:perp} 
& \sup_{0 \le n < n_{\epsilon,\delta,\ell,\alpha}(\lambda\theta)}
\Big\{ \|v_n^\perp(T_{n+1})\|_\infty
+ \big\| v_n(T_n)\|_\infty \Big\} 
\le \epsilon^{\frac 12(1-\alpha)-2\eta},
\\
  \label{eq:apebis}
& 
\sup_{0 \le n < n_{\epsilon,\delta,\ell,\alpha}(\lambda\theta)} \; 
\sup_{t\in [T_n,T_{n+1}]} \| v_n(t) -\sqrt{\epsilon} z_n(t)\|_\infty
 \le \epsilon^{\frac 12 (1-\alpha) -2\eta},
\\
\label{eq:5.4}
&  
\sup_{0 \le n < n_{\epsilon,\delta,\ell,\alpha}(\lambda\theta)} \,
    \Big| x_{n+1} - \Big(x_0 - \frac 34  \sum_{k=0}^{n} 
    \big\langle\bam'_{x_k},v_k(T_{k+1})\big\rangle \Big)  \Big| 
         \le \epsilon^{-\frac 12\alpha  - 3 \eta}\,  \lambda\, T^{-\frac 12},
\end{align}
for any $\epsilon$ small enough.
   \end{theorem}

\proof 
By the recursive definition of $v_n(t)$, see in particular \eqref{4.2} 
and \eqref{gin}, on the event
$\mathcal{B}_{\epsilon,\theta,\eta}^{(1)}$, for $t\le
\epsilon^{-1}\lambda\theta \wedge S_{\delta,\ell,\alpha}$ and $n =
\big[t/T\big]$ we have (where we understand $v_{-1}(0)= m_0
-\bam_{x_0}$) 
   $$
\!\!\!\!\begin{array}{l}
{\displaystyle  
\| v_n(t)\|_\infty 
\le 
\big\| \varphi_{x_n} - g_{t-T_n}^{(x_n)} \varphi_{x_n} \big\|_\infty 
+C_1 \big\|\bam_{x_{n-1}} - \bam_{x_{n}} \big\|_\infty 
+C_1 \big\| v_{n-1}(T_n) \big\|_\infty 
}
\\
{\displaystyle
\phantom{ \| v_n(t)\|_\infty  \le} \vphantom{ \Big[\Big]^{M^{M^M}}}
+ 3\,C_1 \int_{T_n}^t\!ds\, 
 \| v_n(s)\|_\infty^2 \big[1+ \| v_n(s)\|_\infty \big]
+\sqrt{\epsilon T} \epsilon^{-\eta}  
}
\\
\quad \le 
{\displaystyle  
2\sqrt{\epsilon T} \epsilon^{-\eta}
+ C_1(C_0+1) \big\| v_{n-1}(T_n) \big\|_\infty 
+ 3\,C_1 \int_{T_n}^t\!ds\, 
 \| v_n(s)\|_\infty^2 \big[1+ \| v_n(s)\|_\infty \big]
},
\end{array}
   $$
where we used Proposition~\ref{prop2.1} and Lemma~\ref{ape1}, note
$\alpha\in (0,\gamma)$ implies $\epsilon^{-\alpha}<T$.
On the other hand, for $t\in (\epsilon^{-1}\lambda\theta \wedge
S_{\delta,\ell,\alpha}\,,\,\epsilon^{-1}\lambda\theta]$ we clearly have  
$v_n(t) = v_{[S_{\delta,\ell,\alpha}/T]}(S_{\delta,\ell,\alpha})$.
Recalling \eqref{eq:m0}, the proof of \eqref{ape} is now completed 
by a standard bootstrap argument. 

By the recursive definition of $v_n(t)$, Theorem~\ref{specst} and
\eqref{ape}, for $n <  n_{\epsilon,\delta,\ell,\alpha}(\lambda\theta)$,  
on the event $\mc B_{\epsilon,\theta,\eta}$ we have  
   \begin{eqnarray*}
\|v_n^\perp(T_{n+1})\|_\infty & \le & 
C \|\varphi_{x_n} - g_T^{(x_n)}\varphi_{x_n}\|_\infty + 
\epsilon^{\frac 12 -\eta} \\&&  + \; C e^{-\delta_1 T} \|v_n(T_n)\|_\infty^{1/3}
\|v_n(T_n)\|_2^{2/3} + 4\, \epsilon^{1-4\eta} T^2.
   \end{eqnarray*}
Using Lemma~\ref{ape1} and again \eqref{ape},
we can bound the r.h.s.\ above by $\frac 13 \epsilon^{\frac 12(1-\alpha) 
- 2\eta}$. Recalling \eqref{4.2} we have
$$
v_{n+1}(T_{n+1}) =
-\bam_{x_{n+1}} + \bam_{x_n}+ 
\big\langle \phi^{(x_n)}, v_n(T_{n+1}) \big\rangle \, \phi^{(x_n)}  
+ D_n + v_n^\perp(T_{n+1}),
$$
where
$$
D_n := 
\big\langle \Psi_0^{(x_n)}, v_n(T_{n+1}) \big\rangle \, \Psi_0^{(x_n)} 
- \big\langle \phi^{(x_n)}, v_n(T_{n+1}) \big\rangle \, \phi^{(x_n)}. 
$$
From Theorem~\ref{specst} and \eqref{ape} it is straightforward to deduce 
$\|D_n \|_\infty \le \frac 16 \epsilon^{\frac 12(1-\alpha) - 2\eta}$. 
To complete the proof of \eqref{1:perp} it  is then enough to show that
$$
\big\| 
-\bam_{x_{n+1}} + \bam_{x_n}+ 
\big\langle \phi^{(x_n)}, v_n(T_{n+1}) \big\rangle \, \phi^{(x_n)}
\big\|_\infty \le \frac 16 \epsilon^{\frac 12(1-\alpha) - 2\eta},
$$
which follows, by elementary computations, from
Proposition~\ref{prop2.1} and \eqref{ape}, using that there exists 
$C>0$ such that, for any $\epsilon>0$,
\begin{equation}
\label{4/3}
\sup_{|\zeta|\le \alpha a}
\Big[ \int_{-\infty}^\infty\!dx\,\bam_\zeta'(x)^2 - \|\bam_\zeta'\|_2^2 \Big] =
\sup_{|\zeta|\le \alpha a}
\Big[  \frac 43 - \|\bam_\zeta'\|_2^2 \Big] 
\le C \epsilon^{1-\alpha}.
\end{equation}

The bound \eqref{eq:apebis} follows, by \eqref{iequ}, 
from \eqref{ape}, \eqref{1:perp}, and Lemma~\ref{ape1}.

To prove \eqref{eq:5.4}, we first note that, by
Proposition~\ref{prop2.1}, the recursive definition of the center,
and \eqref{ape},  for $n < n_{\epsilon,\delta,\ell,\alpha}(\lambda\theta)$, 
we have
$$
\begin{array}{l}
{\displaystyle
\Big| 
x_{n+1} - x_{n} + \frac 34 \big\langle \bam'_{x_n},v_n(T_{n+1}) \big\rangle 
+\frac 9{16} \big\langle \bam'_{x_n},v_n(T_{n+1}) \big\rangle 
\big\langle \bam''_{x_n},v_n(T_{n+1}) \big\rangle
\Big| 
}
\\
\qquad\qquad\qquad
{\displaystyle\vphantom{\Big\{^{M^M}}
\le C_0 \Big[ \epsilon^{\frac 32 - 6\eta} T^{\frac 32} 
+ 2 \epsilon^{\frac 12 (1-\alpha)} \epsilon^{\frac 12 -2\eta} \sqrt{T} 
\Big].
}
\end{array}
$$
On the other hand, by writing 
$v_n(T_{n+1}) = \big\langle \Psi_0^{(x_n)}, v_n(T_{n+1})\big\rangle
\Psi_0^{(x_n)}+ v_n^\perp(T_{n+1})$ and using \eqref{psi1}, the bound
$\big|\big\langle \bam''_{x_n}, \phi^{(x_n)} \big\rangle \big| \le
\epsilon^{1-\alpha}$ together with \eqref{ape} and \eqref{1:perp} we get
$$
\big| \big\langle \bam'_{x_n},v_n(T_{n+1}) \big\rangle 
\big\langle \bam''_{x_n},v_n(T_{n+1}) \big\rangle
\big| \le \epsilon^{1-\frac \alpha 2 - 4\eta} \sqrt{T}.
$$
Putting together the above estimates we get the bound \eqref{eq:5.4}.
\qed

\section{Recursive equation for the center and stability}
\label{sec:5}
\vskip .2 cm

Let $x_0$ be the center of the initial condition $m_0$ in \eqref{feq},
set $\xi_{0}=x_0$ and 
\begin{eqnarray}
  \label{eq:5.1}
   {\displaystyle  \xi_{n+1}} &:=&
    {\displaystyle 
      x_0 - \frac 34 
      \sum_{k=0}^{n\wedge [S_{\delta,\ell,\alpha}/T]}
      \big\langle \bam'_{x_k},v_k(T_{k+1}) \big\rangle,
      }
    \nn \\
    {\displaystyle
      \sigma_n 
      }
    &:=&
    {\displaystyle
      - \frac 34 
        \sqrt{\epsilon} \big\langle \bam'_{x_n}, z_n(T_{n+1}) \big\rangle
      = - \frac 34 \sqrt{\epsilon} \int_{T_n}^{T_{n+1} }
        \big\langle \bam'_{x_n}, g^{(x_n)}_{T_{n+1}-t} dW(t) \big\rangle,
     \qquad}
     \\ \nn 
     {\displaystyle
       F_n
     }
     &:=&
     {\displaystyle  \frac 34\,\epsilon
        \int_{T_n}^{T_{n+1}} \!dt \,
       \big\langle\bam'_{x_n}, 3 \bam_{x_n} z_n(t)^2 \big\rangle.
       }
   \end{eqnarray}
Notice that, by the bound \eqref{eq:5.4}, $\xi_{n+1}$ is an
approximation to the center $x_{n+1}$ for 
$n< [S_{\delta,\ell,\alpha}/T]$.
Moreover, conditionally on the centers $x_0,x_1,\dots,x_n$, the random
variables $\sigma_0, \ldots, \sigma_n$ are independent Gaussians with mean zero and variance $\frac 34\,
\epsilon T [1+o(1)]$.  The next theorem identifies a recursive
equation satisfied by $\xi_n$.

\begin{theorem}
  \label{t:5.1}
 For each $n< [S_{\delta,\ell,\alpha}/T]$ we have
  \begin{equation}
    \label{eq:5.2}
    \xi_{n+1} -\xi_n = \sigma_{n} + 12 \epsilon T e^{-4\xi_n}
    + F_{n} + R_{n},
  \end{equation}
where the remainder $R_{n}$ can be bounded as follows. There exist
$q,\alpha_0,\eta_0>0$ such that for any $\alpha\in (0,\alpha_0)$, $\eta\in
(0,\eta_0)$, and $\theta\in\bb R_+$ on the event $\mc
B_{\epsilon,\theta,\eta}$ we have
  \begin{equation}
    \label{eq:5.3}
        \sup_{0\le n <  [S_{\delta,\ell,\alpha}/T]} \; |R_n|
        \le \epsilon \lambda^{-1} T \,  \epsilon^q
    \end{equation}
for any $\epsilon$ small enough. Moreover, for each $\theta\in \bb R_+$ 
there exists $q>0$ such that 
  \begin{equation}
   \label{t:5.3}
  \lim_{\epsilon\to 0} \: \bb P \Big( 
  \sup_{0\le n< n_\epsilon(\lambda\theta)}
    \Big|
    \sum_{k=0}^{n} F_k \Big| > \epsilon^q \Big)
    =0.
  \end{equation}
\end{theorem}

We remark that, while the remainder $R_n$ is deterministically small
on the event $\mc B_{\epsilon,\theta,\eta}$, the non-linear term
$F_{n}$ becomes negligible in the limit $\epsilon\to 0$ only in
probability.  This is due to a cancellation in which we exploit a
martingale structure of $F_{n}$. In other words $F_{n}$ gives 
no contribution to the limit equation not because of its magnitude, which
would instead give a finite contribution, but because its
expected value vanishes in the limit. The same mechanism, which
depends on the symmetry of $V$, was
already exploited for the stochastic reaction diffusion equation with
the interface far from the boundary \cite{BBBP1,BBDP,BDP}. 

Before proving Theorem~\ref{t:5.1}, we state a lemma 
that identifies the leading corrections in Lemma~\ref{ape1}
for $t=T$, which will be responsible for the non-linear drift in the
limiting equation \eqref{eq:1.3}.

\begin{lemma}
  \label{t:5.2}
Let $\alpha\in (0,\frac{\gamma}{3})$. Then, for each $\eta>0$, 
   \begin{equation}
   \label{stimphi1}
\lim_{\epsilon\to 0}\, \sup_{|\zeta| < \alpha a} 
\epsilon^{-{(1-\alpha)} +\eta} \,
\Big|
\big\langle \bam_\zeta', \varphi_\zeta - g_T^{(\zeta)}\varphi_\zeta \big\rangle
+ \frac 43 \,  12 \, \epsilon T\, e^{-4 \zeta}  \Big| = 0. 
   \end{equation} 
\end{lemma}

\proof
Recalling \eqref{qa}, we write
   \begin{eqnarray*}
\big\langle  \bam_\zeta', \varphi_\zeta - g_T^{(\zeta)}\varphi_\zeta
\big\rangle
& = &
\big\langle \bam_\zeta', \varphi_\zeta \big\rangle
- e^{-\lambda_0^{(\zeta)}T} 
\big\langle \Psi_0^{(\zeta)}, \varphi_\zeta \big\rangle 
\big\langle \bam_\zeta', \Psi_0^{(\zeta)} \big\rangle 
- \big\langle \bam_\zeta', g_T^{(\zeta,\perp)}\varphi_\zeta \big\rangle
\\ & = &
\big\langle \bam_\zeta', \varphi_\zeta \big\rangle \big[ 1 
- e^{-\lambda_0^{(\zeta)}T} \big\langle \Psi_0^{(\zeta)}, \phi^{(\zeta)} 
\big\rangle \big]
\\ && - \, e^{-\lambda_0^{(\zeta)}T}\big\langle\varphi_\zeta, 
\Psi_0^{(\zeta)} - \phi^{(\zeta)} \big \rangle 
\big\langle \bam_\zeta', \Psi_0^{(\zeta)} \big\rangle 
- \big\langle \bam_\zeta', g_T^{(\zeta,\perp)}\varphi_\zeta \big\rangle.
   \end{eqnarray*}
The last term above is easily bounded by using \eqref{exphi} and
\eqref{gperpin}. Again by \eqref{exphi} and \eqref{psi3} it is easy to 
show, see Lemma~\ref{ape1} for analogous computations, that 
$$
\lim_{\epsilon\to 0}\, \sup_{|\zeta| < \alpha a} 
\epsilon^{-{(1-\alpha)} +\eta} \,
\big\langle \Psi_0^{(\zeta)} - \phi^{(\zeta)}\,,\,\varphi_\zeta \big\rangle 
= 0.
$$
From \eqref{4/3} and since   
   $
\sup_{|\zeta| < \alpha a} \Big| d_\zeta + \frac 12 \Big| 
\le C \epsilon^{\frac 12 (1-\alpha)},
   $
again by \eqref{exphi} we have
$$
\lim_{\epsilon\to 0}\, \sup_{|\zeta| < \alpha a} 
\epsilon^{-\frac 12 (1-\alpha) +\eta} \,
\Big| \big\langle \bam_\zeta' , \varphi_\zeta \big\rangle  
+\frac 23 \Big| = 0.
$$
Finally, by \eqref{psi0} and \eqref{psi3}, 
$$
\lim_{\epsilon\to 0}\, \sup_{|\zeta| < \alpha a} 
\epsilon^{-{(1-\alpha)} +\eta} \, \sup_{|\zeta| < \alpha a} 
\Big\{\big| 1 -  \big\langle \Psi_0^{(\zeta)}, \phi^{(\zeta)} 
\big\rangle \big| 
+ \big| 1-  e^{-\lambda_0^{(\zeta)}T} - 24 \epsilon T e^{-4\zeta}  
\big| \Big\} = 0,
$$
which concludes the proof.
\qed

\medskip
\noindent\textit{Proof of Theorem~\ref{t:5.1}.}
By the recursive definition of $v_n$, we see that \eqref{eq:5.2} holds
with $R_n= -\frac 34 \big[ 
R_n^{(1)} + R_n^{(2)}+ R_n^{(3)}+R_n^{(4)}\big]$ where
\begin{eqnarray*}
{\displaystyle 
R_n^{(1)}
}
&:=& 
{\displaystyle 
\big\langle \bam'_{x_n}, \varphi_{x_n} - g_T^{(x_n)}\varphi_{x_n} \big\rangle
+ \frac 43 \, 12 \epsilon T e^{-4 \xi_n},
}
\\
{\displaystyle 
R_n^{(2)}
}
&:=& 
{\displaystyle 
\big\langle  \bam_{x_n}', g_T^{(x_n)} v_n(T_n) \big\rangle,
}
\\
{\displaystyle 
R_n^{(3)}
}
&:=& 
{\displaystyle - 
\int_{T_n}^{T_{n+1}} \!dt \, \Big\{
\big\langle \bam'_{x_n},  g_{T_{n+1}-t}^{(x_n)}  
\big[ 3 \bam_{x_n} v_n(t)^2  \big]
\big\rangle
-  \epsilon 
\big\langle \bam'_{x_n}, 3 \bam_{x_n} z_n(t)^2 \big\rangle
\Big\},
}
\\
{\displaystyle 
R_n^{(4)}
}
&:=&
{\displaystyle - 
\int_{T_n}^{T_{n+1}} \!dt \, 
\big\langle  \bam'_{x_n},  g_{T_{n+1}-t}^{(x_n)} 
\big[ v_n(t)^3 \big]
\big\rangle.
}
\end{eqnarray*}
The error term $R_n^{(1)}$ is easily bounded by using Lemma~\ref{t:5.2}
and \eqref{eq:5.4}. The bound for the terms $R_n^{(2)}$ and $R_n^{(4)}$ 
follows from Theorem~\ref{apestimate}.  We next bound $R_n^{(3)}$.
By Theorem~\ref{specst} for each $\eta>0$ we have 
$$
\lim_{\epsilon\to 0} \epsilon^{-\frac 12 (1-\alpha) +\eta}
\sup_{t\in[0,T]} 
\sup_{ |\zeta| \le \alpha a }
\big\| \bam'_{\zeta} - g^{(\zeta)}_{t}\bam'_{\zeta} \big\|_1
= 0,
$$
so that, by Theorem~\ref{apestimate} it is enough to prove
\eqref{eq:5.3} for 
$$
\widetilde R_n^{(3)}:= 
\int_{T_n}^{T_{n+1}} \!dt \, 
\big\langle \bam'_{x_n},  \bam_{x_n} 
\big[ v_n(t) - \sqrt{\epsilon}z_n(t)\big]
\big[ v_n(t) + \sqrt{\epsilon}z_n(t)\big]\big\rangle.
$$
We decompose $[T_n,T_{n+1}]=[T_n,T_n+\log^2T] \cup [T_n+\log^2T,T_{n+1}]$ 
and estimate separately the two time integrals. For the first one it
is enough to notice that, by \eqref{ape} and \eqref{eq:apebis}, we have   
$$
\Big| \int_{T_n}^{T_{n}+\log^2T} \!\!\!\!dt \, 
\big\langle \bam'_{x_n},  \bam_{x_n} 
\big[ v_n(t) - \sqrt{\epsilon}z_n(t)\big]
\big[ v_n(t) + \sqrt{\epsilon}z_n(t)\big]\big\rangle
\Big| \le \epsilon^{1-\frac 12 \alpha - 4\eta}\, \sqrt T \, \log^2 T. 
$$
To bound the second integral we write, from the integral equation
for $v_n$, see \eqref{iequ} and the iterative definition of $v_n$, 
 
\begin{eqnarray*}
&& {\displaystyle
v_n(t)-\sqrt{\epsilon} z_n(t) 
}
=
{\displaystyle
   \varphi_{x_n} - e^{-\lambda_0^{(x_n)}(t-T_n)} 
   \big\langle \varphi_{x_n}, \Psi_0^{(x_n)} \big\rangle \Psi_0^{(x_n)}
    -g^{(x_n,\perp)}_{t-T_n}\varphi_{x_n}
}
\\
&& \qquad\qquad\;+\;
 {\displaystyle
    e^{-\lambda_0^{(x_n)}(t-T_n)} 
    \big\langle v_n(T_n), \Psi_0^{(x_n)} \big\rangle \Psi_0^{(x_n)}
+ g^{(x_n,\perp)}_{t-T_n}v_n(T_n) + D_n(t)},
\end{eqnarray*}
where, by Theorem~\ref{apestimate}, $\sup_{t\in[T_n,T_{n+1}]} \|
D_n(t)\|_\infty \le 4 T^2 \epsilon^{1 -4\eta}$.
By the explicit expression \eqref{exphi}, the bound \eqref{vt1} and
Theorem~\ref{specst}, for each $\eta>0$ we have 
$$
\big\langle
\bam_{x_n}' , \big| \varphi_{x_n} - e^{-\lambda_0^{(x_n)}(t-T_n)} 
   \big\langle \varphi_{x_n}, \Psi_0^{(x_n)} \big\rangle \Psi_0^{(x_n)} \big|
\big\rangle \le \epsilon^{1-\alpha -\eta} T.
$$
Since, by the recursive definition of the centers $x_n$, 
$\big\langle \bam'_{x_n},v_n(T_n)\big\rangle = 0$, by \eqref{psi1} and
\eqref{1:perp}, we have 
$$
\big| \big\langle v_n(T_n), \Psi_0^{(x_n)} \big\rangle \big| \le
\epsilon^{1-\alpha-3\eta}. 
$$
Finally, by \eqref{gperpin},  from \eqref{exphi} and \eqref{vt1}, 
since $b=\epsilon^{-\beta}$, there
is $C>0$ such that
$$
\sup_{t\in [T_n+\log^2T,T_{n+1}] } \Big\{ 
\big\| g_{t-T_n}^{(x_n,\perp)}\varphi_{x_n} \big\|_\infty 
+ \big\| g_{t-T_n}^{(x_n,\perp)} v_n(T_n) \big\|_\infty 
\Big\} \le C e^{-\delta_1 \log^2T} \epsilon^{-\frac{\beta}3}.
$$
Putting all the above bounds together and using Theorem~\ref{apestimate} 
to bound $\|v_n(t)+\sqrt{\epsilon}z_n(t)\|_\infty$, we finally get
$$
\Big| \int_{T_{n}+\log^2T}^{T_{n+1}} 
\!\!\!\!dt \, 
\big\langle \bam'_{x_n},  \bam_{x_n} 
\big[ v_n(t) - \sqrt{\epsilon}z_n(t)\big]
\big[ v_n(t) + \sqrt{\epsilon}z_n(t)\big]\big\rangle
\Big| \le \epsilon^{ \frac 32 -\alpha - 6 \eta} \, T^{\frac 72},
$$
which concludes the proof of \eqref{eq:5.3}. 

\smallskip
We next prove \eqref{t:5.3}. By the Doob decomposition, 
   \begin{equation}
\sum_{k=0}^{n-1} F_k = M_n + \sum_{k=0}^{n-1} \gamma_k,
   \end{equation}
where
   \begin{equation}
   \label{gad}
\gamma_k := \bb E\big(F_k\big|\mathcal{F}_{T_k}\big)
   \end{equation}
and $M_n$ is an $\mathcal{F}_{T_n}$-martingale with 
bracket 
   \begin{equation}
   \label{qvm}
\langle M \rangle_n = \sum_{k=0}^{n-1} \Big\{ \bb E\big( F_k^2\big|
\mathcal{F}_{T_k}\big) - \gamma_k^2 \Big\}.
   \end{equation}
Since for $(t,x)\in [T_k,T_{k+1}]\times [-a,b]$
   $$
\bb E \big( z_k(t,x)^2 \big| \mathcal{F}_{T_k}\big) = 
\int_{T_k}^t\! ds\, g_{2(t-s)}^{(x_k)}(x,x),  
   $$
we have
   \begin{eqnarray*}
{\displaystyle
\gamma_k 
}
&=&
{\displaystyle
 - \frac 94 \epsilon \int_{T_k}^{T_{k+1}}\! dt \int_{T_k}^t\! 
ds \int_{-a}^b\!dx\, \bam_{x_k}'(x)\, \bam_{x_k}(x)\, g_{2(t-s)}^{(x_k)}
(x,x)
}
\\
&=&
{\displaystyle
  - \frac 94 \epsilon \int_{0}^{T}\! dt\, (T-t)
\int_{-a}^b\!dx\, \bam_{x_k}'(x)\, \bam_{x_k}(x)\, g_{2t}^{(x_k,\perp)}(x,x)
+r_k,
}
\end{eqnarray*}
where
$$
r_k := - \frac 94 \epsilon\int_{0}^{T}\! dt \, (T-t)
\int_{-a}^b\!dx\, \bam_{x_k}'(x)\, \bam_{x_k}(x)\, 
\exp\big\{-2\lambda_0^{(x_k)}t\big\}\, \Psi_0^{(x_k)}(x)^2.
$$
Since $\big|\big\langle\bam'_{x_k},\bam_{x_k}
\big(\bam'_{x_k}\big)^2\big\rangle \big| \le \epsilon^{\frac 32
(1-\alpha)}$, by \eqref{psi0} and \eqref{psi1} we have that $|r_k|
\le \epsilon^{\frac 32 (1-\alpha)} T^2$.

Recall that  $G^{(\zeta,\perp)}$ has been defined in
\eqref{Gperp}. We claim that
   \begin{equation}
   \label{pippo}
\sup_{|\zeta| < \alpha a} \; \sup_{x\in [-a,b]}
\Big| \int_0^T\!dt\, \frac{T-t}T \, g_{2t}^{(\zeta,\perp)}(x,x)
- \frac 12 G^{(\zeta,\perp)}(x,x) \Big| \le \frac CT.
   \end{equation}
To prove it, we write 
   $$
g_{2t}^{(\zeta,\perp)}(x,x) = \sum_{i=1}^\infty 
\exp\big\{-2\,t\,\lambda_i^{(\zeta)}\big\}\,
\Psi_i^{(\zeta)}(x)^2,
   $$
where $\lambda_i^{(\zeta)}$, resp.\ $\Psi_i^{(\zeta)}$, $i\ge 0$,
are the eigenvalues, resp.\ the eigenfunctions, of $H_\zeta$.
A straightforward computation yields
   $$
\frac 1T \int_0^T\!dt\, (T-t) \, g_{2t}^{(\zeta,\perp)}(x,x)
= \sum_{i=1}^\infty \frac{\Psi_i^{(\zeta)}(x)^2}{2\lambda_i^{(\zeta)}} 
\bigg[1 - \frac{1 - \exp\big\{-2\lambda_i^{(\zeta)}T\big\}}
{2\lambda_i^{(\zeta)}T} \bigg].
   $$
As $G^{(\zeta,\perp)}(x,x) = \sum_{i=1}^\infty
\Psi_i^{(\zeta)}(x)^2/\lambda^{(\zeta)}_i$ the bound \eqref{pippo}
follows from \eqref{gap} and Remark~1 at the end of Section \ref{sec:9}.  
By \eqref{psi4b} and the previous bounds we finally get that there
exists $q>0$ such that
   $$
\sum_{k=0}^{n_\epsilon(\lambda\theta)} \big|\gamma_k\big| \le 
n_\epsilon(\lambda\theta) \sup_{0\le n \le n_\epsilon(\lambda\theta)}
\big|\gamma_n\big| \le \epsilon^q.
   $$

We are left with the bound of the martingale part $M_n$. Given $q>0$, by
Doob's inequality, recalling \eqref{qvm},
   \begin{eqnarray}
   \label{mn}
&& \bb P\Big( \sup_{0\le n \le n_\epsilon (\lambda\theta)} |M_n| \ge
\epsilon^{q} \Big) \le \epsilon^{-2q} \, 
\bb E\Big( \big\langle M \rangle_{n_\epsilon(\lambda \theta)}\Big) 
\nonumber \\ && \qquad \qquad \qquad
\le  \epsilon^{-2q} \sum_{k=0}^{n_\epsilon(\lambda\theta)}
\bb E \big[\bb E \big( F_k ^2 \big| \mc F_{T_k} \big) \big]
\le C^2  \epsilon^{-2q} \big[n_\epsilon(\lambda\theta)+1\big] 
\epsilon^2 T^4, \qquad\quad
   \end{eqnarray}
where we used that there exists $C>0$ such that, for any $\epsilon>0$ and
$k \le n_\epsilon(\lambda\theta)$, we have
   \begin{equation*}
\sqrt{ \bb E \big( F_k ^2 \big| \mc F_{T_k} \big)} 
\le C \epsilon \int_{T_k}^{T_{k+1}}\!dt
\int_{-a}^{b}\! dx\, \bam'_{x_k} (x) 
\sqrt{ \bb E \big( z_k(t,x)^4  \big| \mc F_{T_k} \big)} \le C \epsilon T^2,
   \end{equation*}
which concludes the proof.
\qed

\smallskip
In the following lemma we prove that $\xi_n$ is bounded with
probability close to one. In proving the convergence to the soft wall 
we need such control for $n \le (\epsilon\, T)^{-1}$, while for
the convergence to the hard wall we need that $\xi_n$ grows at most as
$\sqrt\lambda$ for $n\le \lambda(\epsilon\, T)^{-1}$.

   \begin{lemma}
   \label{t:5p1}
For each $\theta\in\bb R_+$ we have
\begin{equation}
  \label{eq:5p1}
  \lim_{L\to\infty}\;
    \varlimsup_{\epsilon\to 0} 
\bb P\big( \sup_{0 \le n \le n_\epsilon(\mu \theta)}
    |\xi_n| >L\sqrt \mu \big) = 0, \qquad \mu=1,\lambda.
\end{equation}
   \end{lemma}

\proof 
Since for $n\ge [S_{\delta,\ell,\alpha}/T]$, by definition
\eqref{eq:5.1}, $\xi_n = \xi_{[S_{\delta,\ell,\alpha}/T]}$,
it is enough to prove the statement for $n <
n_{\epsilon,\delta,\ell,\alpha}(\mu\theta)$.
Recall \eqref{eq:5.2} and let 
   \begin{equation}
   \label{sn}
S_n := \sum_{k=0}^{n-1} \sigma_k, \qquad
A_n := S_n + x_0 + \sum_{k=0}^{n-1} \big[F_k +R_k\big].
   \end{equation}
By \eqref{eq:m0} and Proposition~\ref{prop2.1}, for each $\eta>0$ we
have that, for any $\epsilon$ small enough,
   \begin{equation}
   \label{3volte}
|x_0| \le \epsilon^{\frac 12 -\eta}. 
   \end{equation}
Recalling definition \eqref{eq:5.1}, it is easy to show that there exists
a real $C>0$ such that, for any $\epsilon>0$,
\begin{equation}
\label{acs}
\bb E \big(\sigma_k \big|\mc F_{T_k} \big) =0,
\qquad
\bb E\big(\sigma_k^2 | \mc F_{T_k}\big) \le
C\epsilon T.
\end{equation}  
Given $\theta\in \bb R_+$, an application of Doob inequality then yields
  \begin{equation}
  \label{eq:5p2}
  \lim_{L\to\infty}\;
    \varlimsup_{\epsilon\to 0} 
\bb P\big( \sup_{0 \le n \le n_\epsilon(\mu \theta)}
    |S_n| > L\sqrt \mu \big) = 0.
\end{equation}
By Theorem~\ref{t:5.1}, \eqref{gestperp}, and \eqref{eq:5p2} we have 
\begin{equation}
  \label{eq:6.1.4pre}
  \xi_n=\sum_{k=0}^{n-1} 12 \epsilon T e^{-4\xi_k} + A_n,
\end{equation}
with   
\begin{equation}
  \label{eq:6.1.4}
   \lim_{L\to\infty}\;
    \varlimsup_{\epsilon\to 0}\; \bb P\Big( 
\sup_{0\le n \le n_{\epsilon,\delta,\ell,\alpha}(\mu\theta)}\;
    |A_n| >L\sqrt \mu \Big) = 0.
\end{equation}

Let $\,\overline{\! L\!} := \sup_{n=0,\ldots,n_\epsilon(\mu \theta)}
|A_n|/\sqrt \mu$ and set $L_1:=2 \, (\,\overline{\! L\!}\,+1) + 12 \theta +1
$. To prove \eqref{eq:5p1} we may suppose, and we do now, that
$\epsilon T e^{8L_1\sqrt \mu}\le 1$. We shall then prove that
$\sup_{0\le n \le n_{\epsilon,\delta,\ell,\alpha}(\mu\theta)} |\xi_n|
\le L_1\sqrt \mu$.  Indeed, for any $n\le
n_{\epsilon,\delta,\ell,\alpha}(\mu\theta)$ from \eqref{eq:6.1.4pre}
it is clear that $\xi_n\ge -|A_n|\ge -\,\overline{\! L\!}\,\sqrt \mu$. To
prove the upper bound, by setting $\,\overline{\! n\!}\, (L_1) :=\inf\{n\ge 0:\,
\xi_n >
2L_1\sqrt \mu\} \land n_{\epsilon,\delta,\ell,\alpha}
(\mu\theta)$, we shall prove $\xi_n \le L_1\sqrt \mu$ for 
$n\le \,\overline{\! n\!}\, (L_1)$, which gives $\,\overline{\! n\!}\,(L_1) =
n_{\epsilon,\delta,\ell,\alpha}(\mu\theta)$ and concludes the proof.
Given $n\le \, \overline{\! n\!}\, (L_1)$, let $n_*$ the last up-crossing of
$\sqrt\mu$, namely $n_*=\sup\{ k \le n\,:\, \xi_{k} \le \sqrt\mu
\}$. If $n=n_*$ there is nothing to prove, otherwise from
\eqref{eq:6.1.4pre} we get
  \begin{eqnarray*}
&& \!\!\!\!\!\!\!\ \xi_n = \xi_{n_*}+ A_n -A_{n_*} + 12\epsilon 
T e^{-4\xi_{n_*}}
+ \sum_{k=n_*+1}^{n-1} 12\epsilon T e^{-4\xi_{k}}
\\ && \;\le (2\,\overline{\! L\!}\,+1) \sqrt\mu +  
12\epsilon T e^{8 L_1\sqrt \mu } + 
12 (n-n_*) \epsilon T e^{-4\sqrt \mu}
\\ && \le (2\,\overline{\! L\!}\,+1) \sqrt\mu  + 12 \theta +1 \le L_1\sqrt\mu. 
   \end{eqnarray*}
\qed

\medskip
\noindent
\emph{Proof of Theorem~\ref{t:1}, item (\textit{i}).} \ 
Let us first prove that for each $\theta\in \bb R_+$, $\delta\in
(0,\delta_0)$, $\ell\in (\ell_0,\infty)$ and $\alpha\in (0,1/8)$ we
have
\begin{equation}
  \label{eq:nostop}
\lim_{\epsilon \to 0}\;  
\bb P \big(  S_{\delta,\ell,\alpha} \le  \lambda \epsilon^{-1}\theta  \big) 
= 0.
\end{equation}
Indeed, recalling \eqref{eq:sdl} and \eqref{eq:5.1}, by \eqref{ape}, 
\eqref{eq:5.4}, \eqref{gestperp}, and Lemma~\ref{t:5p1} 
it follows that 
\begin{equation*}
  \lim_{\epsilon \to 0}\;
\bb P \big(  S_{\delta,\ell} \le  \lambda \epsilon^{-1} \theta  \big) 
= 0.
\end{equation*}
By using Proposition~\ref{prop2.1} and again \eqref{ape} we then get 
also \eqref{eq:nostop}. 

We now prove item (\textit{i}) of Theorem~\ref{t:1} with
$X_\epsilon(t) := X \big( m(t\wedge S_{\delta,\ell,\alpha} ) \big)$,
i.e.\ $X_\epsilon(t)$ is the center of the solution to \eqref{ieq}
stopped at $S_{\delta,\ell,\alpha}$. Note $X_\epsilon$ is a continuous
$\mc F_t$-adapted process. Thanks to \eqref{eq:nostop} it is enough to
show that, for each $\theta,\eta>0$,
  \begin{equation}
  \label{eq:item1}
\lim_{\epsilon \to 0}\; \bb P \Big( \sup_{t \le \lambda \epsilon^{-1}
\theta \wedge S_{\delta,\ell,\alpha}} \big\| m(t) -
\bam_{X_\epsilon(t)} \big\|_\infty > \epsilon^{\frac 12 -\eta} \Big)
=0,
\end{equation}
which follows, by taking $\gamma$ small enough, from
Proposition~\ref{prop2.1} and \eqref{ape}.  
\qed

\section{Convergence to the soft wall}
   \label{sec:6}

Recalling that $n_\epsilon(\tau) = [\epsilon^{-1}\tau/T]$,
$T=\epsilon^{-\gamma}$, and that  $\xi_n$ has been defined in
\eqref{eq:5.1}, we define the continuous process $\xi_\epsilon(\tau)$,
$\tau\in\bb R_+$, as the piecewise linear interpolation of $\xi_n$
namely, we set
\begin{equation}
  \label{eq:6.1}
  \xi_\epsilon(\tau) :=  \xi_{n_\epsilon(\tau)} 
  + \big[\tau - \epsilon T n_\epsilon(\tau) \big] 
    \big[\xi_{n_\epsilon(\tau)+1} -\xi_{n_\epsilon(\tau)}\big].
\end{equation}
By \eqref{eq:nostop}, \eqref{eq:5.4}, and \eqref{ape} we have that for
each $\theta\in\bb R_+$ there exists a $q>0$ such that
\begin{equation}
  \label{eq:xcx}
\lim_{\epsilon \to 0}  
\bb P \Big(   
  \sup_{\tau\in [0,\lambda \theta ]} 
  \big| X_\epsilon(\epsilon^{-1} \tau) -\xi_\epsilon(\tau) \big| > 
\epsilon^q \Big) = 0.
\end{equation}

To prove item (\textit{ii}) of Theorem~\ref{t:1} we shall identify the limiting equation satisfied by $\xi_\epsilon$.  To this end we need a
few lemmata.  Recalling the definition \eqref{sn} of $S_n$, we denote by
$S_\epsilon(\tau)$ the continuous process defined, as in
\eqref{eq:6.1}, by the linear interpolation of $S_n$.

The first lemma relies on standard martingale arguments to show the weak
convergence of $S_n$ to a Brownian motion. For completeness we however
present also its proof.

\begin{lemma}
  \label{t:6.-1.1}
As $\epsilon\to 0$, the process $\{S_\epsilon\}$ converges weakly in 
$C(\bb R_+)$ to a Brownian motion with diffusion coefficient 
$\frac 34$.
\end{lemma}

\proof 
Recalling \eqref{acs}, an application of Doob inequality then yields, 
for any $\tau\in \bb R_+$, $\eta>0$
\begin{equation}
  \label{eq:6.1.3b}
\lim_{\delta\to 0} \varlimsup_{\epsilon\to 0} 
     \bb P 
\Big( \sup_{\substack{\tau_1,\tau_2\in [0,\tau] \\ |\tau_2-\tau_1| <\delta}} 
\big| S_\epsilon(\tau_2) -  S_\epsilon(\tau_1) \big| >\eta \Big) = 0.
\end{equation}
Since $S_\epsilon(0)=0$, by \cite[Thm.~8.2]{Billingsley}, 
$\{S_\epsilon\}$ is tight.

Let $S$ be a weak limit of $S_\epsilon$, we shall prove that $S(\tau)$
and $S(\tau)^2 - \frac 34 \tau$ are martingales. By Levy's characterization
theorem we then get the result. By \eqref{acs} we have that, for each
$\tau\in \bb R_+$, $\bb E \big( S_\epsilon(\tau)^2 \big)$ is bounded
uniformly as $\epsilon\to 0$. Let $0\le s_1 \le s_2 \le \cdots \le
s_n\le \tau_1 < \tau_2$, $F$ be a bounded continuous function on $\bb
R^n$, and consider a subsequence, still denoted by $\epsilon$,
converging to zero such that $S_\epsilon\Longrightarrow S$. We then have,
by the boundedness of $F$ and the uniform integrability of
$S_\epsilon(\tau)$,
$$
\begin{array}{l}
{\displaystyle 
\bb E \Big(  
\big[ S(\tau_2) -S(\tau_1) \big] F\big(S(s_1),\cdots,S(s_n)\big)
\Big)
}
\\
\qquad\qquad  
{\displaystyle 
= \lim_{\epsilon\to 0} 
\bb E \Big(  
\big[ S_\epsilon(\tau_2) -S_\epsilon(\tau_1) \big] 
F\big(S_\epsilon(s_1),\cdots,S_\epsilon(s_n)\big)
\Big)
= 0,
}
\end{array}
$$
where we used
$$
S_\epsilon(\tau_2) -S_\epsilon(\tau_1)  = 
\sum_{k=n_\epsilon(\tau_1)}^{n_\epsilon(\tau_2)-1}\sigma_k 
+ \big( \tau_2 -\epsilon T n_\epsilon(\tau_2) \big) \sigma_{n_\epsilon(\tau_2)} 
- \big(\tau_1 -\epsilon T n_\epsilon(\tau_1) \big) \sigma_{n_\epsilon(\tau_1)},
$$
so that $\bb E \big(S_\epsilon(\tau_2) -S_\epsilon(\tau_1) \big| \mc
F_{T_{n_\epsilon(\tau_1)}} \big) =0$. As  $F$, $\tau_1$ and $\tau_2$
 were arbitrary, 
 we get that $S(\tau)$ is a martingale. 

To show the second martingale relationship we first prove the
uniform integrability of $S_\epsilon(\tau)^2$. It is enough to show that,
for each $\tau\in \bb R_+$,
$$
\varlimsup_{\epsilon\to 0} \bb E 
\Big[ \sum_{k=0}^{n_\epsilon(\tau)} \sigma_k \Big]^4 < \infty,
$$
which is proven as follows.  By \eqref{sn}, $S_n$ is a $\mc
F_{T_{n}}$-martingale with quadratic variation $[S]_n
=\sum_{k=0}^{n-1} \sigma_k^2$. By the BDG inequality, see e.g.\
\cite[VII, \S 3]{S}, \eqref{acs},  and the uniform bound $\bb E \big(
\sigma_k^4 | \mc F_{T_k} \big) \le C (\epsilon T)^2$ for some $C>0$,
which follows by a Gaussian computation, we get the above bound.

By \eqref{vt11} we have that, for each $\tau\in\bb R_+$,
$$
\lim_{\epsilon\to 0}\; \sup_{0\le n \le n_\epsilon(\tau)}\;
 \frac 1{\epsilon T} 
\, \Big| \bb E \big( \sigma_n^2 \big| \mc F_{T_n} \big) - 
\frac 34 \epsilon T 
\Big| = 0,
$$
which implies 
$$
\lim_{\epsilon\to 0} 
\bb E \Big( 
\Big[ \sum_{k=n_\epsilon(\tau_1)}^{n_\epsilon(\tau_2)-1} \sigma_k \Big]^2
- \frac 34 (\tau_2-\tau_1) \Big| \mc F_{T_{n_\epsilon(\tau_1)}} \Big)
= 0.
$$
Thanks to the uniform integrability of $S_\epsilon(\tau)^2$, we conclude
that $S(\tau)^2 - \frac 34 \tau$ is a martingale by the same argument 
used to show that $S(\tau)$ is a martingale.
\qed

\medskip

We next show that the process $\{\xi_\epsilon\}_{\epsilon>0}$ is tight.

\begin{lemma}
  \label{t:6.1}
For each sequence $\epsilon\to 0$ the process 
$\xi_\epsilon$ is tight in $C(\bb R_+)$.
\end{lemma}

\proof From \eqref{eq:m0} and Proposition~\ref{t:1}, $\xi_\epsilon(0)\to 0$, so 
 by \cite[Thm.~8.2]{Billingsley}, 
it is enough to show that for each $\tau\in \bb R_+$, $\eta>0$ we have
 \begin{equation}
\label{c2}
\lim_{\delta\to 0} \varlimsup_{\epsilon\to 0} \bb P 
\Big( \sup_{\substack{\tau_1,\tau_2\in [0,\tau] \\ |\tau_2-\tau_1|<\delta}} 
\big| \xi_\epsilon(\tau_2) -  \xi_\epsilon(\tau_1) \big| >\eta \Big) = 0. 
  \end{equation}
By \eqref{eq:6.1} and \eqref{eq:5p1}, to prove
\eqref{c2}  it is enough to show that, for each $L<\infty$
\begin{equation}
  \label{eq:6.3}
\lim_{\delta\to 0} \varlimsup_{\epsilon\to 0} 
     \bb P 
\Big( \sup_{\substack{\tau_1,\tau_2\in [0,\tau] \\ |\tau_2-\tau_1|<\delta}} 
\big| \xi_{n_\epsilon(\tau_2)} - \xi_{n_\epsilon(\tau_1)} \big| >\eta\,,
\sup_{0\le n \le n_\epsilon(\tau)} |\xi_n| \le L \Big) = 0.   
\end{equation}
By Theorem~\ref{t:5.1}, \eqref{gestperp}, and \eqref{eq:nostop}, 
for $\tau_1<\tau_2$,
$$
\xi_{n_\epsilon(\tau_2)}- \xi_{n_\epsilon(\tau_1)} 
= \sum_{k=n_\epsilon(\tau_1)}^{n_\epsilon(\tau_2)-1} 
\big( 12 \, \epsilon T \,  e^{-4 \xi_k} + \sigma_k \big) +
R_\epsilon(\tau_1,\tau_2),
$$ 
where for each $\tau\in \bb R_+$ there exists $q>0$ so that
$$
\lim_{\epsilon\to 0} 
\bb P \Big( \sup_{\tau_1,\tau_2\in [0,\tau]} 
\big|R_\epsilon (\tau_1,\tau_2) \big| > \epsilon^q \Big) = 0.
$$
By \eqref{eq:6.1.3b} it is now straightforward to conclude the
proof of \eqref{eq:6.3}.
\qed

\begin{lemma}
  \label{t:6.2}
For each $\delta>0$, $\theta \in \bb R_+$
$$
\lim_{\epsilon\to 0} \bb P\Big( 
\sup_{s\in[0,\mu\theta]} \; 
\Big| \xi_\epsilon(s) - S_\epsilon(s) - \int_0^s\!du \, 12
\exp\{-4\,\xi_\epsilon(u)\} \Big| > \delta \sqrt\mu \Big) 
=0, \quad\; \mu = 1,\lambda.
$$
\end{lemma}

\medskip
\noindent{\it Remark.} In this section the above lemma is used for $\mu=1$;
we shall use it with $\mu=\lambda$ in proving the convergence to the
hard wall.

\medskip
\noindent\textit{Proof of Lemma~\ref{t:6.2}.}
By Lemma~\ref{t:5p1} it is enough to show that, for each $L>0$,
   \begin{eqnarray}
   \label{pp1}
&& \lim_{\epsilon\to 0} \bb P\Big(\sup_{s\in[0,\mu\theta]} \; 
\Big| \xi_\epsilon(s) - S_\epsilon(s) - \int_0^s\!du \, 12
\exp\{-4\,\xi_\epsilon(u)\} \Big| > \delta \sqrt\mu,
\nonumber \\&&
\quad\qquad\quad\qquad\quad\qquad
 \sup_{0 \le n \le n_\epsilon(\mu \theta)+1} |\xi_n| \le L\sqrt \mu \Big) = 0,
\qquad\qquad \mu = 1,\lambda.\qquad\;
  \end{eqnarray}
Recalling the definition of $\xi_n$ in \eqref{eq:5.1}, the bound
\eqref{ape} and Proposition~\ref{prop2.1} yields $|\xi_{n+1}-\xi_n| \le
C\epsilon^{\frac 12 -\eta}\sqrt T$ for $n \le n_\epsilon(\lambda\theta)$
on a set of probability converging to $1$ as $\epsilon\to 0$
 by \eqref{gestperp}.
By definition \eqref{eq:6.1}, for each $\theta\in\bb R_+$, $\delta>0$, and 
$L>0$ we have 
   \begin{eqnarray}
   \label{pp}
&& \lim_{\epsilon\to 0} \bb P\Big( \sup_{s\in [0,\mu\theta]} \Big|
\sum_{n=0}^{n_\epsilon(s)} \epsilon T
e^{-4\xi_n} - \int_0^s\!du\, e^{-4\xi_\epsilon(u)}\Big|
>\delta \sqrt\mu, 
\nonumber \\&&
\quad\qquad\quad\qquad\quad\qquad
\sup_{0 \le n \le n_\epsilon(\mu \theta)} |\xi_n| \le L\sqrt \mu \Big) = 0,
\qquad\qquad \mu = 1,\lambda,\qquad\qquad
  \end{eqnarray}
as it can be easily shown by the change of variable $u = \epsilon\,t$
in the integral and using $\big| e^{-4\xi_{n+1}} - e^{-4\xi_n}\big| \le
4\,e^{4\max\{|\xi_n|;|\xi_{n+1}|\}}|\xi_{n+1}-\xi_n|$. The proof of
\eqref{pp1} is now completed by using Theorem~\ref{t:5.1}, \eqref{gestperp}, and \eqref{eq:nostop}.
\qed

\medskip
\noindent\emph{Proof of Theorem~\ref{t:1}, item (\textit{ii}).}\
Thanks to \eqref{eq:xcx} it is enough to prove the statement for
$\xi_\epsilon$ in place of $Y_\epsilon$.  Let us denote by
$P_\epsilon$, a probability on $C(\bb R_+)\times C(\bb R_+)$, the law
of the process $(S_\epsilon,\xi_\epsilon)$. By Lemmata~\ref{t:6.-1.1}
and \ref{t:6.1} there exists a subsequence $\epsilon\to 0$ and a
probability $P$ such that $P_\epsilon\Longrightarrow P$. By
\cite[Thm.~III.8.1]{S} there exists a probability space
$\big(\Omega^*,\mc F^*,\bb P^*\big)$ and random elements
$(X^*_\epsilon,Y^*_\epsilon$), $(X^*,Y^*)$ with values in $C(\bb
R_+)\times C(\bb R_+)$ such that the law of
$(X^*_\epsilon,Y^*_\epsilon)$, resp.\ $(X^*,Y^*)$, is $P_\epsilon$,
resp.\ $P$, and $(X^*_\epsilon,Y^*_\epsilon)$ converges to $(X^*,Y^*)$
$\bb P^*$--a.s. Moreover, again by Lemma~\ref{t:6.-1.1}, $X^*$ is a
Brownian motion with diffusion coefficient $\frac 34$. Denoting by
$(x(\cdot),y(\cdot))$ the canonical coordinates in $C(\bb R_+)\times
C(\bb R_+)$, for each $\delta>0$ and $\tau\in\bb R_+$, we have
$$
\begin{array}{l}
{\displaystyle
P \Big( \sup_{s\in[0,\tau]} 
\Big| y(s) - x(s) - \int_0^s\!du \, 12 \exp\{-4y(s)\}  \Big| > \delta
\Big)
}
\\
\qquad
= 
{\displaystyle
\bb P^* \Big( \sup_{s\in[0,\tau]} 
\Big| Y^*(s) - X^*(s) - \int_0^s\!du \, 12 \exp\{-4Y^*(s)\}  \Big| > \delta
\Big) 
}
\\
\qquad=
{\displaystyle
\lim_{\epsilon\to 0}
\bb P^* \Big( \sup_{s\in[0,\tau]} 
\Big| Y^*_\epsilon(s) - X^*_\epsilon(s) - \int_0^s\!du \,  12
\exp\{-4Y^*_\epsilon(s)\}  \Big| > \delta
\Big) 
}
\\
\qquad=
{\displaystyle 
\lim_{\epsilon\to 0}
P_\epsilon \Big( \sup_{s\in[0,\tau]} 
\Big| y (s) - x(s) - \int_0^s\!du \, 12  
\exp\{-4y(s)\}  \Big| > \delta
\Big) 
}
\\
\qquad
=
{\displaystyle
\lim_{\epsilon\to 0}
\bb P \Big( \sup_{s\in[0,\tau]} 
\Big| \xi_\epsilon (s) - S_\epsilon(s) - \int_0^s\!du \,  12 
\exp\{-4\xi_\epsilon (s)\}  \Big| > \delta \Big)
= 0.
}
\end{array}
$$
where we used, in the second step, the $\bb P^*$--a.s convergence of
$(X^*_\epsilon,Y^*_\epsilon)$ to $(X^*,Y^*)$ and, in the last step,
Lemma~\ref{t:6.2} with $\mu=1$. As $\delta$ and $\tau$ were arbitrary
it follows that any limit point solves \eqref{eq:1.3}. In fact this
also prove existence of a weak solution to \eqref{eq:1.3}. 
 Since the real function $y\to 12
e^{-4y}$ is locally Lipschitz, by \cite[Thm.\ 5.2.5]{KS} there is
path-wise uniqueness of \eqref{eq:1.3}. By \cite[Cor.\ 5.3.23]{KS} it
follows there is a strong solution to \eqref{eq:1.3} which is unique
in the sense of probability law. We then conclude that $Y_\epsilon$
weakly converges to the unique strong solution of \eqref{eq:1.3} \qed
  
\section{Convergence to the hard wall} 
\label{sec:7} 

To prove item (\textit{iii}) of Theorem~\ref{t:1}, we first state and prove an
analogous result for the diffusive scaling of the stochastic equation
\eqref{eq:1.3}. To simplify the notation we introduce a probabilistic
model not related with the one introduced in Section~\ref{sec:2} and
denote by $t$ the macroscopic time variable.  Let $B$ be a Brownian
motion on some filtered probability space $\(\Omega,\mc F,\mc F_t,\mc
P\)$ and $\gamma$ a positive parameter that will eventually diverge.
We suppose given a sequence of $\mc F_t$-adapted continuous processes
$B_\gamma$ such that $B_\gamma(0)=0$ and satisfying that for each $T\ge 0$, 

\begin{equation}
  \label{eq:bg-b}
  \mc P \Big( \lim_{\gamma\to \infty} \sup_{t\in [0,T]}
  \big| B_\gamma(t) - B(t) \big| =0 \Big) = 1.
\end{equation}
We consider the sequence of processes that solve the equation
   \begin{equation}
   \label{1p1}
Y_\gamma(t) = 
\gamma \int_0^t\!ds \, \big[ Y_\gamma(s)\big]_- + B_\gamma(t),
   \end{equation}
where $[Y]_- = \max\{0,-Y\}$ is the negative part of $Y$.  We shall
prove that $Y_{\gamma}$ converges to a Brownian motion reflected at
the origin.  The precise statement is the following.

   \begin{theorem}
   \label{1p2}
Let 
   \begin{equation}
   \label{1p3}
Y(t) := B(t) + \sup_{s\in[0,t]} \{-B(s)\}.
   \end{equation}
Then, for any $T\ge0$, 
$$
\mc P \Big( \lim_{\gamma\to \infty} \sup_{t\in [0,T]}
  \big| Y_\gamma(t) - Y(t) \big| =0 \Big) = 1.
$$

   \end{theorem}
   
Note that, by e.g.\ \cite[Thm.~6.17]{KS}, $Y$ has the law of a
Brownian motion reflected at the origin. 

\proof 
Let 
\begin{equation}
  \label{eq:rg}
  r_\gamma(T) := \sup_{t\in[0,T]}\,\sup_{\gamma'> \gamma} 
  \big| B_{\gamma'}(t) - B_{\gamma}(t) \big|
\end{equation}
and note that by \eqref{eq:bg-b}, for each $T\in \bb R_+$ we have 
$r_\gamma(T)\to 0$ $\mc P$-a.s. as $\gamma\to \infty$. We claim that
for $\gamma_1<\gamma_2$, $t\in [0,T]$, we have
\begin{equation}
  \label{eq:mon}
  Y_{\gamma_2} (t) \ge  Y_{\gamma_1} (t) - 2 r_{\gamma_1} (T).
\qquad  
\end{equation}
Indeed, if $Y_{\gamma_2} (t) \ge  Y_{\gamma_1} (t)$ there is nothing
to prove, otherwise let $\tau = \sup\{ s\in [0,t] \, : \, 
 Y_{\gamma_2} (s) \ge Y_{\gamma_1} (s) \}$ which exists because 
$Y_{\gamma_2} (0) = Y_{\gamma_1} (0)$. By definition,  
$Y_{\gamma_2} (s) \le  Y_{\gamma_1} (s)$ for $s\in [\tau, t]$; by
writing the equation \eqref{1p1} in this interval and using the
monotonicity of $x\mapsto [x]_-$ the bound \eqref{eq:mon} follows
easily.  

We next claim that 
\begin{equation}
  \label{eq:mon2}
  Y_{\gamma}(t) \le B_\gamma (t) + \sup_{s\in[0,t]} \{-B_\gamma (s)\}
  =: w_\gamma(t).
\end{equation}
This can be proved as follows. We first note that $w_\gamma\ge 0$. 
Let $t\ge 0$, if $Y_{\gamma}(t) \le 0$ there is nothing to prove,
otherwise, setting $\tau = \sup\{s\in [0,t] : Y_{\gamma}(s) = 0\}$ we have:
   $$
Y_{\gamma}(t) = 
Y_{\gamma}(t) - Y_{\gamma}(\tau) = B_\gamma(t) - B_\gamma(\tau) 
+ \int_\tau^t\!ds\, 
\gamma \big[ Y_{\gamma}(s) \big]_- = B_\gamma(t) - B_\gamma(\tau) 
\le w_\gamma(t),
   $$
where we used that $\big[Y_{\gamma}(s)\big]_- = 0$ for $s\in
[\tau,t]$. 

Let $Z(t) := \varliminf_{\gamma\to \infty} Y_\gamma(t)$. By
\eqref{eq:bg-b} and \eqref{eq:mon2} we have $Z(t) \le Y(t)$.  It is
easy to show, by \eqref{eq:mon}, that $\mc P$-a.s.\
$\lim_{\gamma\to\infty} Y_\gamma(t) = Z(t)$.  To complete the proof of
the theorem we shall prove: $Z$ is a.s.\ continuous, $Z \ge 0$, there
exists a continuous increasing process $\ell$ so that $Z=B+ \ell$ and
$\int_0^\infty \! d\ell(t) \, Z(t) = 0$. Then from the Skorohod Lemma,
see e.g.\ \cite[Lemma~6.14]{KS}, it follows $Z=Y$.

For $f\in C(\bb R_+)$, $\delta>0$, and $T>0$, we let $\omega_{\delta,T}(f)$ be
the modulus of continuity of the function $f$ on $[0,T]$, i.e.\  
   $$
\omega_{\delta,T}(f) := \sup_{\substack{s,t\in [0,T] \\ |t-s|<\delta}}
|f(t) - f(s)|.
   $$
We first show the \emph{a priori} bound:
   \begin{equation}
   \label{ygs1}
\inf_{t\in [0,T]} Y_\gamma(t) \ge - 2\, \omega_{\delta,T}(B_\gamma) -4\,
e^{-\delta \gamma} \sup_{t\in [0,T]} |B_\gamma(t)|.
   \end{equation}
Indeed, pick $\tau \in [0,T]$ such that $\inf_{t\in [0,T]} Y_\gamma(t) = 
Y_\gamma(\tau)$. If $Y_\gamma (\tau) = 0$ there is nothing to prove,
otherwise let $\sigma = \sup\{t\in [0,\tau] : Y_\gamma(t) = 0\}$.
For $t\in [\sigma,\tau]$ we can integrate the equation \eqref{1p1}
getting:
   \begin{eqnarray*}
Y_\gamma(\tau) & = & B_\gamma(\tau) - B_\gamma(\sigma) 
- \int_\sigma^\tau\!ds\, 
\gamma e^{-(\tau-s)\gamma}\, [B_\gamma(s) - B_\gamma(\sigma)] \\
& = & e^{-(\tau-\sigma)\gamma}\, [B_\gamma(\tau)-B_\gamma(\sigma)] 
+ \int_\sigma^\tau\!ds\, 
\gamma e^{-(\tau-s) \gamma}  [B_\gamma(\tau) - B_\gamma(s)] \\
& = & e^{-(\tau-\sigma)\gamma}\, [B_\gamma(\tau)-B_\gamma(\sigma)] + 
\int_\sigma^{\sigma\vee (\tau-\delta)} \!ds\, 
\gamma e^{-(\tau-s)\gamma}\, [B_\gamma(\tau) - B_\gamma(s)] \\
&& + \, \int_{\sigma\vee (\tau-\delta)}^\tau \!ds\, 
\gamma e^{-(\tau-s)\gamma}\, [B_\gamma(\tau) - B_\gamma(s)] \\
& \ge & - 4 \, e^{-\delta\gamma} \sup_{t\in [0,T]} |B_\gamma(t)| - 
2\, \omega_{\delta,T}(B_\gamma).
   \end{eqnarray*}

We next bound the modulus of continuity of $Y_\gamma$. We claim that
   \begin{equation}
   \label{ygs2}
\omega_{\delta,T}(Y_\gamma) \le 8 \Big[ \omega_{\delta,T}(B_\gamma) + 
e^{-\delta\gamma}\sup_{t\in [0,T]}|B_\gamma(t)| \Big].
  \end{equation}
Let us fix $t,s\in [0,T]$ with $|t-s|<\delta$. We consider first the case 
in which $Y_\gamma(u)\le 0$ for any $u \in [s,t]$. Solving equation \eqref{1p1}
in this time interval, we get
   \begin{eqnarray*}
Y_\gamma(t) - Y_\gamma(s) 
& = & \big( e^{-(t-s)\gamma} - 1 \big) Y_\gamma(s)   
+ B_\gamma(t) - B_\gamma(s) 
\\ && 
- \int_s^t\!du\, \gamma e^{-(t-u)\gamma}\,  
\big[B_\gamma(u)-B)_\gamma(s)\big],
   \end{eqnarray*}
so that, by \eqref{ygs1}, 
   \begin{equation}
   \label{ygs3}
\big|Y_\gamma(t) - Y_\gamma(s)\big| 
\le \big|Y_\gamma(s)\big| + 2\,\omega_{\delta,T}(B_\gamma) 
\le 4 \Big[ \omega_{\delta,T}(B_\gamma) 
+  e^{-\delta \gamma}\sup_{t\in [0,T]}|B_\gamma(t)| \Big].
   \end{equation}
The case in which $Y_\gamma(u)\ge 0$ for any $u \in [s,t]$ we
clearly have $\big|Y_\gamma(t) - Y_\gamma(s)\big| \le
\omega_{\delta,T}(B_\gamma)$.  
The other cases can be reduced to the previous ones. 
We discuss only the case $Y_\gamma(s)<0$, $Y_\gamma(t)<0$. Let
$\sigma = \inf\{u>s : Y_\gamma(u) = 0\}$ and $\tau = \sup\{u<t :
Y_\gamma(u) = 0\}$. We then write $\big|Y_\gamma(t) - Y_\gamma(s)\big|
= \big|Y_\gamma(t) - Y_\gamma(\tau)\big| + \big|Y_\gamma(\sigma) -
Y_\gamma(s)\big|$ and use the bound \eqref{ygs3} in the 
intervals $[s,\sigma]$ and $[\tau,t]$ to get \eqref{ygs2}.

By taking the limit as $\gamma \to \infty$ in \eqref{ygs2} we get that
the limiting process $Z$ is continuous.  Let ${\overline{\!
Y\!}}_\gamma\,(t) :=\inf_{\gamma'\ge \gamma} Y_{\gamma'}(t)$ so that
${\overline{\! Y\!}}_\gamma\, (t)\uparrow Z(t)$. By the continuity of
$Z$, the previous convergence is in fact uniform for $t$ on compacts.
By using \eqref{eq:mon} we get that $Y_\gamma(t)- {\overline{\!
Y\!}}_\gamma(t)$ converges, $\mc P$-a.s., to zero uniformly for $t\in
[0,T]$.  Hence $Y_\gamma$ converges to $Z$ uniformly on compacts.

To show that $Z\ge 0$ we note that
   $$
\int_0^t\!ds\, \big[ Y_\gamma(s) \big]_- 
= \frac 1\gamma \big[ Y_\gamma(t) - B_\gamma(t) \big],
   $$
which, by taking first the limit $\gamma\to \infty$ and
then $t\to\infty$, implies $\int_0^\infty\!ds\, 
\big[ Z(s) \big]_- = 0$, whence $Z\ge 0$ by the continuity of $Z$. 

Let us introduce the increasing process 
   $$
\ell_\gamma(t) := \int_0^t\!ds\, \gamma 
\big[ Y_\gamma(s)\big]_- = Y_\gamma(t) - B_\gamma(t).
   $$
By the convergence of $Y_\gamma$ to the continuous process $Z$, 
   $$
\ell(t) := \lim_{\gamma\to\infty} \ell_\gamma(t) = Z(t) - B(t)
   $$
is a continuous increasing process. In particular the
Lebesgue-Stieltjes measure $d\ell_\gamma$ weakly converges to $d\ell$
as $\gamma\to\infty$. To finally show $\int_0^\infty\! d\ell(t)\, Z(t)
= 0$ we note that the support of the measure $d\ell_\gamma$ is
a subset of $\{t \ge 0 : Y_\gamma(t) \le 0\}$. By the uniform
convergence of $Y_\gamma$ to $Z$ and the weak convergence of
$d\ell_\gamma$ to $d\ell$, we have, for each $T\in \bb R_+$, 
   $$
\int_0^T \! d\ell(t)\, Z(t) =  
\lim_{\gamma\to\infty} \int_0^T \! d\ell_\gamma(t)\, 
Y_\gamma(t) \le 0,
   $$
and we are done since $Z\ge 0$. \qed

\medskip
Given $\gamma>0$, let $X_\gamma$ be the solution of the equation
   \begin{equation}
   \label{1p4bis}
X_\gamma(t) = 
\gamma \int_0^t\!ds \,  12  \exp\{-4\gamma X_\gamma(s)\} 
+ B_\gamma(t).
   \end{equation}
Note that if $Y(\tau)$ solves \eqref{eq:1.3} then $X_\lambda(t) :=
\lambda^{-1/2} Y(\lambda t)$ solves \eqref{1p4bis} in law with $\gamma
=\sqrt\lambda$ and $B_\gamma$ a Brownian motion for each 
$\gamma$.

   \begin{corollary}
   \label{1p4}
As $\gamma \to \infty$ the process $X_\gamma$ converges $\mc P$ almost surely
to the continuous process $Y$ defined by \eqref{1p3}.
   \end{corollary}

\proof For given $\delta>0$, set $c_{\delta,\gamma} : = 
12 \gamma e^{-4\gamma\delta}$ and define the continuous process 
$Z_{\delta,\gamma}$ as
   \begin{equation}
   \label{1p5}
Z_{\delta,\gamma}(t) := \delta + B_\gamma(t) + c_{\delta,\gamma} t + 
\sup_{s\in [0,t]} \big[-B_\gamma(s) -c_{\delta,\gamma} s \big].
   \end{equation}
Note that $Z_{\delta,\gamma}(0) = \delta$. Recall that $Y_\gamma$ is
the solution of \eqref{1p1}. By arguing as in the proof of
Theorem \ref{1p2} the following comparison holds. For each $\delta>0$ and
$\gamma >1$, we have, $\mc P$ almost surely, 
  \begin{equation}
  \label{1p6}
Y_\gamma \le X_\gamma \le Z_{\delta,\gamma},
   \end{equation}
from which, by using Theorem \ref{1p2}, the statement follows by taking 
first the limit as $\gamma\to \infty $ and then as $\delta\to 0$.
\qed 

\medskip
We are now ready to conclude the proof of our main result. We next
denote by $\theta$ the macroscopic time variable and recall $\lambda =
\log\epsilon^{-1}$. Recalling $\xi_\epsilon$ is defined in
\eqref{eq:6.1}, let $\zeta_\epsilon$ be the continuous process defined
as $\zeta_\epsilon(\theta) := {\lambda}^{-1/2}\xi_\epsilon ({\lambda}
\theta)$.

   \begin{lemma}
   \label{t:8.3}
Let 
   \begin{equation}
   \label{bl}
B_\epsilon(\theta) := \zeta_\epsilon(\theta) - 
\sqrt{\lambda}\int_0^\theta\!ds \, 12 \,
\exp\{ -4 \sqrt\lambda\, \zeta_\epsilon(s)\}.
  \end{equation}
The process $B_\epsilon$ weakly converges in $C(\bb R_+)$ 
to a Brownian motion with diffusion coefficient $\frac 34$.
   \end{lemma}

\proof Recalling $S_\epsilon$ is the linear interpolation of the
sequence $S_n$ defined in \eqref{sn}, let $\,\overline{\!
S\!}_\epsilon(\theta) := {\lambda}^{-1/2} S_\epsilon ({\lambda}
\theta)$. By arguing exactly as in Lemma \ref{t:6.-1.1}, one shows
that the process $\,\overline{\! S\!}_\epsilon$ weakly converge in
$C(\bb R_+)$ to a Brownian motion with diffusion $\frac 34$.
Moreover, by Lemma~\ref{t:6.2} with $\mu=\lambda$, for each
$\delta>0$, $\theta\in \bb R_+$, we have
   \begin{equation*}
\lim_{\epsilon\to 0} \bb P\Big( 
\sup_{s\in[0,\theta]} \; 
\Big| \zeta_\epsilon(s) - \overline{\! S\!}_\epsilon(s) -
\sqrt{\lambda} \int_0^s\!du\, 12  \, 
e^{-4 \sqrt\lambda\, \zeta_\epsilon(u)} \Big| > \delta \Big) = 0,
   \end{equation*}
which concludes the proof.
\qed

\medskip
\noindent\emph{Proof of Theorem~\ref{t:1}, item (\textit{iii}).}\ Thanks to
\eqref{eq:xcx} it is enough to prove the statement for
$\zeta_\epsilon$ in place of $Z_\epsilon$.  By Lemma~\ref{t:8.3} and
\cite[Thm.~III.8.1]{S} there exists a probability space
$\big(\Omega^*,\mc F^*,\bb P^*\big)$ and random elements $B^*$,
$B^*_\epsilon$, with values in $C(\bb R_+)$ such that $B^*$ is a
Brownian motion with diffusion coefficient $\frac 34$, the law of
$B^*_\epsilon$ equals the one of $B_\epsilon$ defined in \eqref{bl},
and $B^*_\epsilon$ converges, $\bb P^*$ almost surely, to $B^*$. We
now define $\zeta^*_\epsilon$ as the solution of the equation
   $$
\zeta^*_\epsilon(\theta) = B^*_\epsilon(\theta) - 
\sqrt\lambda \int_0^\theta\!ds \, 12 \,
\exp\{ -4 \sqrt\lambda\, \zeta^*_\epsilon(s)\}.
  $$
By uniqueness of its solution, the law of $\zeta^*_\epsilon$ equals
the one of $\zeta_\epsilon$. By Corollary~\ref{1p4}
$\zeta^*_\epsilon(\theta)$ converges, $\bb P^*$ almost surely, to
$B^*(\theta) + \sup_{s \le \theta} \{-B^*(s)\}$, whose law is that of
a Brownian motion with diffusion coefficient $\frac 34$ reflected at the
origin.  
\qed

   \section{Spectral Analysis}
   \label{sec:9}

In this section we prove Theorem~\ref{specst}.
To keep the notation simple we shall  define the operator
\begin{equation}
\label{op}
H = - \frac 12 \Delta + V''(\bam), \qquad \bam(x) := \varth(x),
\end{equation}
acting on $L_2([-a,b])$ with Dirichlet boundary conditions. We denote
by $\lambda_0 < \lambda_1 < \ldots < \lambda_i < \ldots$, resp.\
$\Psi_i$ (recall $\Psi_0$ is chosen positive), $i\ge 0$, the
eigenvalues, resp.\ the eigenfunctions, of $H$ and by $g_t :=
\exp\{-tH\}$ the corresponding semigroup. The operators $g^\perp_t$
and $G^\perp$ are defined as in \eqref{gperp} and \eqref{Gperp}.

By standard techniques it is not difficult to compute the Green operator  
$G=H^{-1}$ for the quartic double well potential $V$ in \eqref{3.2a} obtaining that its integral kernel is given by:
   \begin{equation}
   \label{Gxy}
G(x,y)=\frac{2\bam'(x)\bam'(y)}{h(b)+h(a)}
\begin{cases} \big[h(x)+h(a)\big]\, \big[h(b)-h(y)\big] \;\;\text{if } 
-a \le x \le y \le b 
\\ \big[h(y)+h(a)\big]\, \big[ h(b)-h(x)\big] \;\;\text{if } 
-a \le y < x \le b  
\end{cases}
   \end{equation}
where, recalling \eqref{eq:2.1},
   \begin{equation}
   \label{h}
h(x) := h_0(x) = \frac 38 x + \frac 38 \frac{\bam(x)}{\bam'(x)} + 
\frac14 \frac{\bam(x)}{\bam'(x)^2}.
   \end{equation}

\noindent{\it Notation warning.}
In the sequel we will denote by $C$ a generic positive constant,
independent of $a,b$, whose numerical value may change from line to
line and from one side to the other in an inequality.
 
\smallskip
We first obtain some rougher estimates by following the approach in
\cite[Lemma 2.1]{Chen} where analogous bounds are proven in the case
of Neumann boundary conditions.

   \begin{lemma}
   \label{t:chen}
There exists $K>0$ and $a_*>0$ such that, for any $b\ge a\ge a_*$, 
   \begin{eqnarray}
   \label{l0ab0}
&& 0 \le \lambda_0 \le  K\, e^{-4a},
\vphantom{K_{K_K}} 
\\ 
\label{psab0} 
&&
\langle \Psi_0, \bam' \rangle \ge \frac 1{K},
\\ 
&&
\label{gapab0}
\lambda_1 - \lambda_0  \ge  \frac 1{K},
\\ 
&&
\label{gapab1}
\|\Psi_0\|_\infty + \|\Psi_0'\|_\infty \le K.
   \end{eqnarray}
   \end{lemma}

\noindent\emph{Sketch of the proof.}\/

\smallskip
\noindent\emph{Step 1.}
An elementary computation shows that, for each $f \in C^2_0([-a,b])$,
   $$
\langle f, H f \rangle = \frac 12 \int_{-a}^b\!dx\, \bam'(x)^2
\left[\frac{d}{dx}\, \frac{f(x)}{\bam'(x)}\right]^2 \ge 0,
   $$
which in particular implies $\lambda_0 \ge 0$. On the other hand, by 
using $G\bam'$ as test function in the variational characterization 
of the smallest eigenvalue,
   \begin{equation}
   \label{raq}
\lambda_0\,\le \frac{\langle G \bam', H G \bam'\rangle} {\|G\bam'\|^2_2}
= \frac{\langle \bam', G \bam'\rangle}{\| G \bam'\|^2_2}.
   \end{equation}
From \eqref{Gxy} we now get
   \begin{equation}
   \label{Gbam'}
G \bam'(x) = 2 h(a)\, \|\bam'\|_2^2 \, \bam'(x) + 2 \bam'(x) B(x) + A(x),
   \end{equation}
where
   \begin{align}
   \label{A}
A(x) \, & = \, \frac{-2\bam'(x)\, [h(x)+h(a)]}{h(b)+h(a)} 
\int_{-a}^b\!dy\, \bam'(y)^2 \, [h(y)+h(a)],
\\ \label{B}
B(x)\,& = \,\int_{-a}^x\!dy\, \bam'(y)^2 h(y) + 
h(x) \int_x^b\!dy\, \bam'(y)^2.
   \end{align}
Then:
   \begin{align}
   \label {Gbam}
\langle \bam', G \bam'\rangle \, = \; & 2 h(a)\, \|\bam'\|_2^4 
+ 2 \langle \bam', \bam'B \rangle + \langle\bam', A \rangle,
\\ \nn
\|G \bam'\|_2^2 \, = \; & 4 h^2(a)\, \|\bam'\|_2^6 + 
8 h(a) \, \|\bam'\|_2^2\, \langle \bam', \bam'B \rangle 
+ 4 \| \bam'B\|_2^2 
\\ \label{GG2bam}
& +\, 4 h(a) \, \|\bam'\|_2^2 \, \langle  \bam',A \rangle + 
4 \langle \bam'B, A\rangle + \|A\|_2^2.
   \end{align}
 From \eqref{A} and \eqref{ch} we get
   \begin{equation}
   \label{cA}
\|A\|_{\infty}\le C\, b \, \frac{h^2(a)}{\,h(b)^{1/2}}, \qquad 
\|A\|_2 \le  C\, b \, \frac{h^2(a)}{\,h(b)^{1/2}},
   \end{equation}
and, from  \eqref{B} and \eqref{ch}, 
   \begin{equation}
   \label{pwe}
|\bam'(x)\,B(x)|\le \bam'(x)(x+a) + C\, e^{-2x},
   \end{equation}
so that, after integrating,
   \begin{equation}
   \label{cB}
\langle \bam',\bam'B\rangle\le C\,a, \qquad
\|\bam'B\|_2^2 \le C\, e^{4a}.
    \end{equation}
Substituting \eqref{Gbam} and \eqref{GG2bam} in the last quotient in 
\eqref{raq}, after estimating the terms with the aid of \eqref{cA}, 
\eqref{cB} and  \eqref{h}, the bound \eqref{l0ab0} follows.

\smallskip
\noindent\emph{Step 2.} 
Let $\psi$ be an eigenfunction associated to an eigenvalue $\lambda\le
1/2$ and choose a real $\ell_0$ such that $\inf_{|x|\ge \ell_0} 
V''(\bam(x)) \ge 3/2$. By a comparison principle, we get:
   \begin{equation}
   \label{2.8}
\begin{array}{lcl}
{\displaystyle
|\psi(x)| 
}
&\le & 
{\displaystyle
|\psi(\ell)|\,
\frac{\varsh(\sqrt{2}(b-x))}{\varsh(\sqrt{2}(b-\ell))} 
\quad\quad\quad\quad\;\, \forall\, \ell_0 \le \ell \le x \le b, 
}
\\
{\displaystyle
|\psi(x)| 
}
&\le & 
{\displaystyle
|\psi(-\ell)|\,
\frac{\varsh(\sqrt{2}(x+a))}{\varsh(\sqrt{2}(-\ell+a))} 
\quad\quad\quad \forall\, -a\le x \le - \ell \le - \ell_0. 
}
\end{array}
   \end{equation}
Since $\psi$ is normalized there exist reals $\ell_\pm$,
 $\ell_+\in [\ell_0,\ell_0+1]$  and $\ell_-\in [-\ell_0-1,-\ell_0]$ 

such that $|\psi(\ell_\pm)|\le 1$. Hence, for any $b\ge a > \ell_0+1$, 
   \begin{equation}
\label{2.9}
|\psi(x)|\le  C \exp\{-\sqrt{2} |x|\}
\quad\quad\quad\quad \forall \, |x| \ge \ell_0+1.
   \end{equation}

\noindent\emph{Step 3.}
By \eqref{2.9} there exist reals $\ell^*,a^*>0$ such that
$\int_{-\ell^*}^{\ell^*} \!dx \, \Psi_0(x)^2 \ge 1/2$ for any
$a>a^*$. Since $\lambda_0$ is uniformly bounded by \eqref{l0ab0}, by
the Harnack inequality applied to the equation $[H
-\lambda_0]\Psi_0=0$ in the interval $[-\ell^*-1,\ell^*+1]$ we get
that, for any $b\ge a \ge a^*$, we have
   \begin{equation*}
\inf_{|x|\le \ell^*} \Psi_0 (x)
\ge C \sup_{|x|\le \ell^*} \Psi_0 (x) \ge 
C \Big[\frac{1}{2\ell^*} \int_{-\ell^*}^{\ell^*} \!dx \, \Psi_0(x)^2 
\Big]^{1/2} \ge \frac{C}{2\sqrt{\ell^*}}.
   \end{equation*}
The above bound and $\bam'(x) = \varch(x)^{-2}$ yields \eqref{psab0}.

\smallskip
\noindent\emph{Step 4.}\/ 
We can assume $\lambda_1 \le 1/2$. As well known, the corresponding 
eigenfunction $\Psi_1$ has a unique zero $x_0$ in the open interval 
$(-a,b)$; moreover, by \eqref{2.8}, $|x_0|< \ell_0$. Integration by 
parts and $H \bam' =0$ yields
   \begin{equation*}
\lambda_1 \ge \lambda_1 \Big| \int_{x_0}^{b} \!dx \, 
\Psi_1(x) \, \bam'(x) \Big| = \frac 12 \Big| (\Psi_1'\bam')(x_0)
- (\Psi_1'\bam')(b) \Big| \ge \frac 12 \Big| (\Psi_1'\bam')(x_0) \Big|
   \end{equation*}
since 
$\mathrm{sgn}[(\Psi_1)'(x_0)\, (\Psi_1)'(b)]=-1$.

By the same argument as in Step 3, we have that either 
$\int_{x_0}^{\ell^*} \!dx \,  \Psi_1(x)^2  \ge 1/4$ or
$\int_{-\ell^*}^{x_0} \!dx \,  \Psi_1(x)^2  \ge 1/4$. By using
the Hopf maximum principle we then deduce a lower bound on 
$|\Psi_1'(x_0)|$ which is uniform in $b\ge a \ge a^*$. The
estimate \eqref{gapab0} follows.

\smallskip
\noindent\emph{Step 5.}\/ A uniform bound for $\|\Psi_0\|_\infty$
follows from \eqref{2.9} and a comparison argument in the
interval $[-\ell_0-1,\ell_0+1]$. Finally, since $H\Psi_0 = \lambda_0\Psi_0$ 
and $|V''(\bam)| \le 2$, we have $|\Psi_0''(x)| \le C \Psi_0(x)$. The
bound \eqref{gapab1} follows.
\qed

\medskip
\noindent\emph{Proof of Theorem~\ref{specst}.} We observe that, 
given $\alpha\in (0,1)$ it is equivalent to prove 
\eqref{gin}--\eqref{psi4b} for the operator $H$ with
   \begin{equation}
   \label{abz}
a = \frac 14 \log\epsilon^{-1} + \zeta, \qquad
b= \epsilon^{-\beta} - \zeta, \qquad 
|\zeta| \le \frac 14 \alpha \log\epsilon^{-1},
   \end{equation}
and that Lemma~\ref{t:chen} clearly holds for these values of $a$ and $b$.
  
\medskip
\noindent{\it Proof of \eqref{gin}.} 
By the Feynman-Kac formula, see e.g.\ \cite[Theorem \ 2.3]{Fr}, we have 
that, for any $f\in C([-a,b])$, $t>0$, and $x\in (-a,b)$,
   \begin{equation}   
   \label{fk}
\big(g_t f\big) (x)  = \bb E \Big( f(B_t^{(x)}) \id_{\{\tau_x >t\}} 
\exp\Big\{\int_0^t\!ds\, V''(\bam(B_s^{(x)}))\Big\}\Big),
   \end{equation}
where $\{B_t^{(x)},\, t\ge 0\}$ is a Brownian motion starting at $x$
and $\tau_x := \inf\{t \ge 0: B_t^{(x)} \notin (-a,b)\}$. 
The above representation permits to compare $g_t$ with the semigroup 
$\exp\{-t \, \overline{\!H\!}_0 \}$, defined on the whole line $\bb R$. 
For the latter the analogous estimate has been proved in 
\cite[Prop. A.8]{B}, whence
   $$
\big\vert (g_t f) (x)\big\vert\le \big(g_t |f|\big) (x) \le 
\big(\exp\{-t \, 
\overline{\! H\!}_0 \} |f|\big) (x) \le C\Vert f\Vert_{\infty}. 
   $$

\medskip
\noindent{\it Proof of \eqref{gap}.} 
It is a restatement of  \eqref{gapab0}.

\medskip
\noindent{\it Proof of \eqref{gperpin}.}  
We will use an interpolation inequality, see \cite[Lemma 5.1]{FM},
that holds for each $F \in C^1([-a,b])$ such that $F(a)=F(b)=0$,
   \begin{equation}
   \label{fmi}
\|F\|_{\infty}^3 \le \frac 32 \|\nabla F\|_\infty \|F\|_2^2.
   \end{equation}
Recalling $p_t^0$ denotes the heat semigroup with zero boundary conditions 
at the endpoints of $[-a,b]$, we have:
   $$
\nabla g_t f = \nabla p^0_t f - \int_0^t\!ds\,
\nabla p^0_{t-s} V''(\bam) g_s f. 
   $$
Since $\|\nabla p^0_t f\|_\infty \le C t^{-\frac 12} \|f\|_\infty$, by
\eqref{gin} and the above identity we conclude that $\|\nabla g_t f
\|_\infty \le C \sqrt t \|f\|_\infty$ for any $t\ge 1$. By choosing
$F=g_t^\perp f$ in \eqref{fmi}, the estimate \eqref{gperpin} follows
from \eqref{gperp}, \eqref{gapab1} and \eqref{gap}.  
\qed

\medskip
To prove the estimates \eqref{psi0}--\eqref{psi4b}, we will use the
Kellogg method, see e.g.\ \cite{vladimirov}, to obtain successive
approximations of the eigenvalues and eigenvectors by iterations of
the Green operator $G$ applied to the function $\phi(x) :=
\phi^{(0)}(x) = \|\bam'\|_2^{-1} \bam'(x)$, $x\in [-a,b]$.  Let $f_0
:= \phi$, $f_1 := Gf_0$, $f_2:=G f_1$ and $e_1:= f_1/ \|f_1\|_2$, 
$e_2:= f_2/ \|f_2\|_2$ be their $L_2$-normalizations. Also, let
   \begin{equation}
   \label{Rcab}
\mu := \frac{\|f_1\|_2} {\|f_2\|_{2}},\quad
R^2 :=\sup_{x\in[-a,b]} \int_{-a}^b\!dy \, G(x,y)^2,
\quad c:=\langle \Psi_0, \phi \rangle.
   \end{equation}
Then, by \cite[\S 28.1]{vladimirov}, we have the estimates
   \begin{equation}
   \label{ukellog}
\begin{array}{l} 
{\displaystyle 0\le \mu - \lambda_0  
\le \frac{\lambda_0}{2}\,
\bigg[\frac{\lambda_0}{\lambda_1} \bigg]^2 
\frac{ 1 - c^2} {c^2},}
\\ {\displaystyle \vphantom{\bigg\{^{K^{K^K}}}
\big\|\Psi_0 - e_1 \big\|_2 \le
\frac{\lambda_0}{\lambda_1} 
\frac{\sqrt{1 - c^2}}{c},}
\\ {\displaystyle \vphantom{\bigg\{^{K^{K^K}}}
\big\| \Psi_0 - e_2 \big\|_\infty
\le R\, \la_1\, 
\bigg[ \frac{\lambda_0}{\la_1} \bigg]^2
\frac{\sqrt{ 1 - c^2}}{c}.} 
\end{array}
   \end{equation}

To use the above estimates, we will need expressions for $e_i$,
$i=1,2$, and $\mu$. They are given in terms of the following
formulae. From \eqref{Gxy} and \eqref{Gbam'}, we have:
   \begin{equation}
   \label{GGbam}
G^2 \bam' (x) = P(x) + U(x),
   \end{equation}
where
   \begin{align}
\nn P(x) \, & = \, 4 h^2(a) \, \|\bam'\|_2^4\, \bam'(x) +
4h(a) \, \|\bam'\|_2^2 \, \bam'(x) B(x) +  2 G(\bam'B)(x),
\\ \label{U}
U(x) \, & = \, 2 h(a)\, \|\bam'\|_2^2 \, A(x) + GA (x).
   \end{align}
Also, 
   \begin{align}
\nn \|G^2\bam'\|_2^2 \, = \; & 16h^4(a) \, \|\bam'\|_2^{10}
+ 32 h^3(a) \, \|\bam'\|_2^6 \, \langle\bam',\bam'B\rangle +
16h^2(a)\, \|\bam'\|_2^4 \, \|\bam'B\|_2^2 
\\ \nn & 
+\, 16 h^2(a) \, \|\bam'\|_2^4 \, \langle G \bam ',\bam'B \rangle 
+ 16 h(a)\, \|\bam'\|_2^2\, \langle \bam'B,G(\bam'B)\rangle 
\\ \label{GGbam2} 
& +\, 4\|G(\bam'B)\|_2^2 + \|U\|_2^2 + 2 \langle P,U \rangle.
   \end{align}

We finally remark that, by \eqref{psab0}, $c = \|\bam'\|_2^{-1}
\langle \Psi_0, \bam' \rangle$ is uniformly bounded from below by some
positive constant.

\medskip
\noindent{\it Proof of \eqref{psi0}}.
By \eqref{l0ab0} and \eqref{abz} we have that, for each $\eta>0$,
   \begin{equation}
   \label{lagz}
\lim_{\epsilon \to 0} \, \sup_{|\zeta| < \frac 14 \alpha \log\epsilon^{-1}} 
\epsilon^{-(1-\alpha)+\eta}\, \la_0 = 0.
  \end{equation} 
From \eqref{ukellog}, \eqref{gap}, and \eqref{lagz}, to prove \eqref{psi0} 
it is enough to show that
   \begin{equation}
   \label{muz}
\lim_{\epsilon \to 0} \, \sup_{|\zeta| < \frac 14 \alpha \log\epsilon^{-1}}
\epsilon^{-\frac32 (1-\alpha)+\eta}  \:
\big |\mu - 24\, \epsilon \, e^{-4\zeta} \big| = 0. 
   \end{equation}
From \eqref{GG2bam}, the estimates \eqref{cA}, \eqref{cB}, and 
\eqref{abz}, it follows that
   \begin{equation}
   \label{D1}
\|G\bam'\|_2 = 2 h(a) \|\bam'\|_2^3 \, (1+\Delta_1), \qquad
\lim_{\epsilon \to 0} \, \sup_{|\zeta| < \frac 14 \alpha \log\epsilon^{-1}} 
\epsilon^{-\frac 12 (1-\alpha)+\eta} \Delta_1 = 0.
   \end {equation}
Analogously, from \eqref{GGbam2} and the estimate $\|G(\bam'\,B)\|_2^2\le 
C\, a^4\, h^2(a)$ (that follows from \eqref{Gxy} and \eqref{pwe}), together 
with \eqref{cA} and \eqref{cB}, 
   \begin{equation}
   \label{D2}
\|G^2\bam'\|_2=4h^2(a)\|\bam'\|_2^5\, (1+\Delta_2), \qquad 
\lim_{\epsilon \to 0} \, \sup_{|\zeta| < \frac 14 \alpha \log\epsilon^{-1}} 
\epsilon^{-\frac 12 (1-\alpha)+\eta} \Delta_2 = 0.
   \end {equation}
Substitution of the previous expressions in the definition of $\mu$ 
yields
   \begin{equation}
   \label{Mu}
\mu = \frac{1}{2h(a)\,\|\bam'\|_2^2} \, \frac{1+\Delta_1}{1+\Delta_2},
   \end{equation}
from which \eqref{muz} follows since, by \eqref{abz}, $ 
\epsilon \, e^{-4\zeta} = e^{-4a}$ and, by \eqref{h},
\begin{equation}
   \label{Mu'}
\lim_{\epsilon \to 0} \, \sup_{|\zeta| < \frac 14 \alpha \log\epsilon^{-1}}
\epsilon^{-\frac32 (1-\alpha)+\eta} \bigg| \frac{1}{2h(a)\,\|\bam'\|_2^2}
- 24\, e^{-4a} \bigg| = 0.
   \end{equation}
   
\medskip
\noindent{\it Proof of \eqref{psi2}.}
By \eqref{Rcab} and \eqref{Gxy} we have $R^2 = \sup_{x\in[-a,b]} \,
G(x,x) \le C\, h(a)$.  From \eqref{ukellog}, \eqref{lagz}, and
\eqref{gap}, to prove \eqref{psi2} it is then enough to show that,
for each $\eta>0$,
   \begin{equation}
   \label{e2phi}
\lim_{\epsilon \to 0} \, \sup_{|\zeta| < \frac 14 \alpha \log\epsilon^{-1}} 
\epsilon^{-\frac 12 (1-\alpha)+\eta} \, \big\| e_2 - \phi \big\|_\infty = 0. 
   \end{equation}
From the definition of $e_2$, \eqref{GGbam}, \eqref{U}, and \eqref{D2},
we have:
   \begin{multline}
   \label{e2psi}
e_2(x) - \phi(x) =  \frac{G^2\bam'(x)}{\|G^2\bam'\|_2}
- \phi(x) \\ \quad =
\bigg(\frac{\bam'(x)B(x)}{h(a) \|\bam'\|_2^3} +
\frac{G(\bam'B)(x)}{2h^2(a)\|\bam'\|_2^5} +
\frac{U(x)}{4h^2(a)\|\bam'\|_2^5}\bigg)\frac{1}{1+\Delta_2}
- \frac{\Delta_2\,\phi(x)}{1+\Delta_2}.
   \end{multline}
Now, by \eqref{cA}, \eqref{pwe}, \eqref{U}, and using the definition
\eqref{Gxy}, it is easy to show that:
   \begin{align}
&\lim_{\epsilon \to 0} \, \sup_{|\zeta| < \frac 14 \alpha \log\epsilon^{-1}}
\epsilon^{-\frac 12 (1-\alpha)+\eta}\; \frac{\|\bam'B\|_\infty}{h(a)}
 = 0, \nn \\ \label{Ginf} &
\lim_{\epsilon \to 0} \, \sup_{|\zeta| < \frac 14 \alpha \log\epsilon^{-1}}
\epsilon^{- (1-\alpha)+\eta}\; 
\frac{\|U\|_\infty + \|G(\bam'B)\|_\infty}{h^2(a)} = 0,
   \end{align} 
from which \eqref{e2phi} follows.

\medskip
\noindent{\it Proof of \eqref{psi1}.}
For any $\ell<a$ we have
   $$
\big\|\Psi_0 - \phi\big\|_1 \le 2\ell \,\big\|\Psi_0 - \phi\big\|_\infty
+ \int_{[-a,b]\setminus[-\ell,\ell]}\!dx\, 
\big(\Psi_0(x) + \phi(x)\big).  
   $$
Then, by \eqref{psi2}, \eqref{2.9}, and recalling $\phi(x)\le e^{-2|x|}$, 
we get \eqref{psi1} by choosing e.g.\ $\ell = \log^4\epsilon^{-1}$.

\medskip
\noindent{\it Proof of \eqref{psi3}.} 
To prove \eqref{psi3} recall that, from \eqref{ukellog}, \eqref{lagz},
and \eqref{gap}, it is sufficient to show that, for each $\eta>0$,
   \begin{equation*}
\lim_{\epsilon \to 0} \, \sup_{|\zeta| < \frac 14 \alpha \log\epsilon^{-1}} 
\epsilon^{- (1-\alpha)+\eta} \; \big\langle 
\big|e_2 - \phi \big|, \phi \big\rangle =0.
   \end{equation*}
Substituting in $\langle \big| e_2\, - \phi \big|, \phi \big\rangle$ 
the expression \eqref{e2psi}, since $\|\phi\|_1\le C$, the limit above
follows from \eqref{Ginf} and the first estimate in \eqref{cB}.

\medskip
\noindent{\it Proof of \eqref{psi4b}.}
From the definition \eqref{Gxy}, \eqref{eq:2.1}, and recalling that
$\lambda_0$ is the first eigenvalue of $H$, we have:
   \begin{equation}
   \label{gg0}
G^\perp(x,x) = 2 \bam'(x)^2 \big[h(x)+h(a)\big]\,
\bigg[1-\frac{h(x)+h(a)}{h(b)+h(a)}\bigg] - \frac{\Psi_0^2(x)}{\la_0}.
   \end{equation}
Since $h(x)\bam'(x)^2 \le C$ (see \eqref{ch}),  
   \begin{equation}
   \label{gg1}
\int_{-a}^b\!dx\, \bam'(x)^3 \, |\bam(x)| \,
\frac{\big[h(x)+h(a)\big]^2}{h(b)+h(a)}
\le C \, \frac{\sqrt{h(b)}+h^2(a)}{h(b)+h(a)}.
   \end{equation}
We next notice that, by \eqref{psi3} and the definition \eqref{Rcab}
of $c$, for each $\eta>0$, 
   $$
\lim_{\epsilon \to 0} \, \sup_{|\zeta| < \frac 14 \alpha \log\epsilon^{-1}} 
\epsilon^{- \frac 12 (1-\alpha)+\eta} \;\sqrt{1-c^2} = 0,
   $$
so that, by \eqref{ukellog}, \eqref{lagz}, and \eqref{gap}, 
   \begin{equation}
   \label{gg2}
\lim_{\epsilon \to 0} \, \sup_{|\zeta| < \frac 14 \alpha \log\epsilon^{-1}} 
\epsilon^{- \frac 12 (1-\alpha)+\eta} 
\int_{-a}^b\!dx\, \bam'(x) \, |\bam(x)|\, \bigg|
\frac{\Psi_0^2(x)}{\la_0} - \frac{e_1^2(x)}{\mu}\bigg| = 0.
   \end{equation}
On the other hand, from \eqref{D1}, \eqref{D2}, and \eqref{Mu},
   \begin{equation}
   \label{e2q}
\frac{e_1(x)^2}{\mu} = \frac{1+\Delta_3}{2h(a)\|\bam'\|_2^4}
\big[G\bam'(x)\big]^2, \qquad
\lim_{\epsilon \to 0} \, \sup_{|\zeta| < \frac 14 \alpha \log\epsilon^{-1}} 
\epsilon^{-\frac 12 (1-\alpha)+\eta} \Delta_3 = 0.
   \end{equation}
Taking the square in \eqref{Gbam'} and substituting into \eqref{e2q},
from \eqref{cA}, \eqref{pwe}, \eqref{cB} and \eqref{e2q}, it follows that 
   $$
\frac{e_1(x)^2}{\mu} = 2 h(a) (1+\Delta_3)\bam'(x)^2 + 
\frac{4B(x)\bam'(x)^2}{\|\bam'\|_2^2} + W(x),
   $$
with 
   $$
\lim_{\epsilon \to 0} \, \sup_{|\zeta| < \frac 14 \alpha \log\epsilon^{-1}}
\epsilon^{-\frac 12 (1-\alpha)+\eta}\; \big\|(\bam')^2\,\bam\,W\big\|_1=0.
   $$
By \eqref{gg0}, \eqref{gg1}, \eqref{gg2}, and the above limit, we are reduced
to prove that, for each $\eta>0$,
   \begin{equation}
   \label{PI}
\lim_{\epsilon \to 0} \, \sup_{|\zeta| < \frac 14 \alpha \log\epsilon^{-1}}
\epsilon^{-\frac 12 (1-\alpha)+\eta}\; Q(\zeta,\epsilon) = 0, 
   \end{equation}
where
   \begin{equation}
Q(\zeta,\epsilon) \, := \, \bigg| \int_{-a}^b\!dx\, \bam'(x)^3\, \bam(x)\, 
\bigg(2\,h(x) - \frac{4B(x)}{\|\bam'\|_2^2} - 2\Delta_3h(a)\bigg) \bigg|.
   \end{equation}
Now, since $(\bam')^3\bam$ is an odd function, we have 
   \begin{equation}
   \label{P1}
\bigg| \int_{-a}^b\!dx\,\bam'(x)^3 \bam(x)\bigg| =
\bigg| \int_{a}^b\!dx\,\bam'(x)^3 \bam(x)\bigg| \le C\, e^{-4a}.
    \end{equation}
From \eqref{B}, 
   $$
h(x)-\frac{2B(x)}{\|\bam'\|_2^2} = h(x) \bigg[ 1 - 2
\int_{x}^b\! dy\, \frac{\bam'(y)^2}{\|\bam'\|_2^2} \bigg] - 
2 \int_{-a}^x\!dy\, \frac{\bam'(y)^2}{\|\bam'\|_2^2} \, h(y).
   $$
Since $\bam(x) = \varth(x)$,
   $$
2 \int_{x}^b\! dy\, \frac{\bam'(y)^2}{\|\bam'\|_2^2} = 
\frac 32 \int_x^\infty\!dy\, \bam'(y)^2 + D(x) = 
1 + \frac{\bam^3(x)-3 \bam(x)}2 + D(x),
   $$
with $|D(x)|\le C e^{-4 a}$. Then, recalling $h(x)\bam'(x)^2 \le C$,
   \begin{align}
   \label{P2}
& \bigg| \int_{-a}^b\!dx\,\bam'(x)^3 \bam(x)\,h(x) \bigg[ 1 - 2 
\int_{x}^b\! dy\, \frac{\bam'(y)^2}{\|\bam'\|_2^2} \bigg] \bigg| 
\nn \\ & \qquad \, \le \, C \, e^{-4 a} + \bigg| \int_{-a}^b\!dx\, \bam'(x)^3
\bam(x)\,h(x) \, \frac{\bam^3(x)-3 \bam(x)}2 \bigg|
\nn \\ & \qquad \, \le \, C \, e^{-2 a},
   \end{align}   
where we have used that the integrand in the last integral is an odd
function, see \eqref{h}. Finally, observing that $-6(\bam')^3 \bam =
[(1-\bam^2)^3]'$, integration by parts in the remaining integral yields
   \begin{align}
   \label{P3}
\nn & \bigg| \int_{-a}^b\!dx\, \bam'(x)^3\bam(x) \int_{-a}^x\!dy\,
 \bam'(y)^2 \, h(y)\bigg| \; \le \; 
\bigg| \frac{\big(1-\bam^2(b)\big)^3}{6} \int_{-a}^b\!dy\, \bam'(y)^2 \, h(y)\bigg|
\\ & \qquad\qquad\qquad + \, \bigg| \int_{-a}^b\!dx\, 
\frac{\big(1-\bam^2(x)\big)^3}{6} \bam'(x)^2 \, h(x) \bigg| \, \le \, C \, e^{-6a}.  
   \end{align}
To estimate the last integral we have used again $h(x)\bam'(x)^2 \le C$
and that the integrand is an odd function.  From \eqref{P1}, \eqref{P2}, 
and \eqref{P3} the limit \eqref{PI} follows.
\qed
   
\medskip
\noindent{\it Remark 1.}
Proceeding as in the proof of \eqref{psi4b}, from \eqref{psi0}, 
\eqref{psi2}, and \eqref{ukellog} it can be shown that 
$\sup_{x\in[-a,b]}G^\perp(x,x)<\infty$.

\medskip
\noindent{\it Remark 2.}
From the previous computations,  it follows  that $G^\perp(x,y)$
 converges pointwise, as $\epsilon \to 0$, to the kernel of the generalized 
Green function $\,\overline{\! G\!}$ which inverts $ \overline{\!H\!}_0$
on the subspace orthogonal to $\bam'$. This kernel is
  \begin{equation}
  \label{eq:p1}
  \overline{\! G\!}\,(x,y):= \begin{cases}{\displaystyle
\frac 34 \bam'(x) \bam'(y) 
\Big[u(x) + u(-y) +\frac 5{12}\Big]}
 & \textrm{if $x\le y$} 
\\ \\ {\displaystyle
\frac 34 \bam'(x) \bam'(y) 
\Big[u(-x) + u(y) +\frac 5{12}\Big]}
 & \textrm{if $x>y$} 
\end{cases}
\end{equation}
where
\begin{equation}
  \label{eq:p2}
  u(x):=  \frac 1{24} e^{4x} + \frac 13 e^{2x} + \frac 12 x - \frac 38.
\end{equation}
This expression has been obtained in \cite[Prop.~3.3]{PS} where
however the constant $\frac 52$ should read $\frac{5}{12}$.

\appendix
\section{Fluctuations of a localized interface}
\label{s:A}

In this section we sketch the proof of Theorem~\ref{t:10}, which
describes the asymptotic behavior of the interface when $a=b=
\frac 14 \log\epsilon^{-1}$, by pointing out the relevant 
differences w.r.t.\ the case $a=\frac 14\log\epsilon^{-1}$,
$b\gg a$. We then explain how to get the uniformity w.r.t.\
the initial condition. 

\smallskip
\noindent
\emph{Sketch of the proof of Theorem~\ref{t:10}.} Fix $\tau_0>0$.
Throughout this section we denote by $m(t;m_0)$, $t\in 
[0,\epsilon^{-1}\tau_0]$, the solution to \eqref{feq} with 
$a=b=\frac 14 \log\epsilon^{-1}$, to emphasize its dependence on the
initial condition $m_0\in \mc X_\epsilon$. Accordingly, we let
$X(m_0)$, resp.\ $X(t;m_0)$, be the center of $m_0$, resp.\
$m(t\wedge S_{\delta,\ell,\alpha};m_0)$,
see \eqref{eq:sdla}. Recalling the set $\mc N_\eta^\epsilon(z)$
is defined in the statement of the theorem, for each $L>0$ we define
$N_\eta^{\epsilon,L} := \bigcup_{z\in [-L,L]} \mc N_\eta^\epsilon(z)$.
The iterative scheme of Section~\ref{sec:3} is repeated with no changes
in the present setting.

\smallskip
\noindent
\emph{Step 1. Spectral analysis.} We claim that Theorem~\ref{specst}
holds with the only change that the asymptotic \eqref{psi0} for
the smallest eigenvalue has to be replaced by 
\begin{equation}
\label{psi0u}
\lim_{\epsilon \to 0} \, \sup_{|\zeta| < \alpha a}
\epsilon^{-\frac 32 (1-\alpha)+\eta}  \:
\big |\lambda_0^{(\zeta)} - 48\, \epsilon \varch(4\zeta) \big| = 0.
\end{equation}
As in Section~\ref{sec:9}, we fix the center at the origin and study 
the operator \eqref{op} in the interval $[-\ell-\zeta,\ell-\zeta]$.
The asymptotic of the eigenvalue $\lambda_0^{(\zeta)}$ can be obtained
as in \eqref{ukellog}. The asymptotic of $\mu$, as defined \eqref{Rcab},
is obtained as follows. Instead of \eqref{Gbam'} we here decompose
\begin{eqnarray*}
G\bam'(x) & = &  2\, \|\bam'\|_2^2 \, \frac{ h(\ell+\zeta)\, 
h(\ell-\zeta)}{h(\ell+\zeta)+h(\ell-\zeta)}\, \bam'(x)
\\&& + \, \frac{2 \,h(\ell-\zeta)\, \bam'(x)}
{h(\ell+\zeta)+h(\ell-\zeta)} 
\bigg[\int_{-\ell-\zeta}^x\!dy\, \bam'(y)^2 h(y) + 
h(x) \int_x^{\ell+\zeta}\!dy\, \bam'(y)^2\bigg]
\\&& - \, \frac{2 \,h(\ell+\zeta)\, \bam'(x)}
{h(\ell+\zeta)+h(\ell-\zeta)}
\bigg[h(x) \int_{-\ell-\zeta}^x\!dy\, \bam'(y)^2 + 
\int_x^{\ell+\zeta}\!dy\, \bam'(y)^2 \, h(y) \bigg]
\\ && - \, \frac{2\,h(x)\,\bam'(x) }{h(\ell+\zeta)+h(\ell-\zeta)} 
\int_{\ell-\zeta}^{\ell+\zeta}\!dy\, \bam'(y)^2 \, h(y)
\end{eqnarray*}
and get
\begin{eqnarray*}
\|G\bam'\|_2   & = & 2 \,
\frac{h(\ell+\zeta)h(\ell-\zeta)}{h(\ell+\zeta)+h(\ell-\zeta)}
\, \|\bam'\|_2^3 \, (1+\widetilde\Delta_1), \\
\|G^2\bam'\|_2 & = & 4 
\bigg[\frac{h(\ell+\zeta)h(\ell-\zeta)}{h(\ell+\zeta)+h(\ell-\zeta)}
\bigg]^2 \, \|\bam'\|_2^5 \, (1+\widetilde\Delta_2),
\end{eqnarray*}
where $\widetilde\Delta_1$ and $\widetilde\Delta_2$ satisfy the
estimates stated in \eqref{D1} and \eqref{D2} for $\Delta_1$ and 
$\Delta_2$. The bound \eqref{psi0u} now follows by direct computations,
see \eqref{Mu} and \eqref{Mu'}.

\smallskip
\noindent
\emph{Step 2. A priori bounds and recursive equation for the center.}
The a priori bounds of Section~\ref{sec:4} depend only on $b\ge a$
and therefore hold also in the present setting. Moreover, there exists
$\eta_1>0$ such that the following holds. For each $L>0$ and
$\eta\in [0,\eta_1]$ there exists $\eta_0>0$ such that the bounds 
stated in Theorem~\ref{apestimate} hold for $\eta\in (0,\eta_0)$ 
uniformly w.r.t.\ $m_0$ in the set $\mc N_{\eta'}^{\epsilon,L}$,
$\eta'\in [0,\eta_1]$. 

The key estimate \eqref{stimphi1} in Lemma~\ref{t:5.2} for the identification of the nonlinear drift is here replaced by
   \begin{equation}
   \label{stimphi1u}
\lim_{\epsilon\to 0}\, \sup_{|\zeta| < \alpha a} 
\epsilon^{-{(1-\alpha)} +\eta} \,
\Big|
\big\langle \bam_\zeta', \varphi_\zeta - g_T^{(\zeta)}\varphi_\zeta \big\rangle
+ \frac 43 \,  24 \, \epsilon T\, \varsh(4\zeta)  \Big| = 0, 
   \end{equation} 
which is proven as follows. Recalling \eqref{exphi}, we have
\begin{eqnarray*}
&& \sup_{|\zeta| < \alpha a} \Big| d_\zeta + \frac 12 \Big| 
\le C \epsilon^{\frac 12 (1-\alpha)} 
\\ && \sup_{|\zeta| < \alpha a} \Big| c_\zeta - 1 \Big| 
\le C \epsilon^{\frac 12 (1-\alpha)}
\\ && \bigg|  \frac{h(\ell+\zeta) - 
h(\ell-\zeta)}{h(\ell+\zeta)+h(\ell-\zeta)} -
\frac 12 + \frac 12 \varth(4\zeta) \bigg|  \le
C \epsilon^{\frac 12 (1-\alpha)},
\end{eqnarray*}
whence
$$
\lim_{\epsilon\to 0}\, \sup_{|\zeta| < \alpha a} 
\epsilon^{-\frac 12 (1-\alpha) +\eta} \,
\Big| \big\langle \bam_\zeta' , \varphi_\zeta \big\rangle  
+\frac 23  \varth(4\zeta) \Big| = 0.
$$
In view of this bound and \eqref{psi0u}, we can repeat the computations
in Lemma~\ref{t:5.2} and get \eqref{stimphi1u}.

Let $\xi_n$ and $\sigma_n$ be defined as in \eqref{eq:5.1}. We
emphasize that $\xi_0 = x_0 = X(m_0)$ so that the whole sequence
$\xi_n$ depends on the initial condition $m_0$. By using 
\eqref{stimphi1u} and following the same steps as in 
Theorem~\ref{t:5.1}, it is easy to prove its analogue in
the present setting with a uniform control on $m_0\in 
\mc N_\eta^{\epsilon,L}$, $\eta\in [0,\eta_1]$. 
Set $b(x) := - 24\varsh(4x)$, then 
\begin{equation}
\label{eq:5.2u}
\xi_{n+1} -\xi_n = \sigma_{n} + \epsilon T b(\xi_n) + \Theta_{n},
\end{equation}
where, for each $L\in \bb R_+$, there exists $q>0$ such that 
  \begin{equation}
   \label{t:5.3u}
\lim_{\epsilon\to 0} \; \sup_{m_0\in \mc N_\eta^{\epsilon,L}}\;
\bb P \Big(\sup_{0\le n< \epsilon^{-1}\tau_0/T}
\Big| \sum_{k=0}^{n} \Theta_k \Big| > \epsilon^q \Big) = 0.
  \end{equation}
Moreover, by the same argument as in Lemma~\ref{t:5p1}, the above
statement implies that, for each $L\in\bb R_+$ we have
\begin{equation}
  \label{eq:5p1u}
  \lim_{K\to\infty}\; \varlimsup_{\epsilon\to 0} \; 
\sup_{m_0\in \mc N_\eta^{\epsilon,L}}\; 
\bb P\Big( \sup_{0 \le n \le n_\epsilon(\tau_0)} |\xi_n| >K \Big) = 0,
\end{equation}
which yields, see the end of Section~\ref{sec:5},  
\begin{equation}
  \label{eq:nostopu}
\lim_{\epsilon \to 0}\; \sup_{m_0\in \mc N_\eta^{\epsilon,L}}\;
\bb P \big(  S_{\delta,\ell,\alpha} \le  \epsilon^{-1}\tau_0  \big) 
= 0
\end{equation}
and
\begin{equation}
  \label{eq:1.2u}
\lim_{\epsilon\to 0} \; \sup_{m_0\in \mc N_\eta^{\epsilon,L}}\;
\bb P \Big( \sup_{t\in [0,\epsilon^{-1}\tau_0]} \big\|
m(t;m_0)-\bam_{X(t;m_0)}\big\|_{\infty} > 
\epsilon^{\frac 12 -\eta} \Big) = 0.
\end{equation}

\smallskip
\noindent
\emph{Step 3. A coupling argument.}
By \eqref{eq:1.2u}, the uniform convergence \eqref{eq:1.31} follows
once we show there exists $\eta_1>0$ such that for each 
$\eta\in[0,\eta_1]$, $L > 0$, and each uniformly continuos and bounded
function $F\, : \, C([0,\tau_0]; \mc X) \to \bb R$, we have 
\begin{equation}
\label{egu}
\lim_{\epsilon\to 0} \; \sup_{z_0 \in [-L,L]} \;
\sup_{m_0\in \mc N_\eta^\epsilon(z_0)}\;
\big| \bb E\,F\big(\bam_{Y(\cdot;m_0)}\big) -
E \,F\big( \bam_{\Xi^{z_0}(\cdot)}\big)\big|\,=\, 0,
\end{equation}
where $Y(\tau;m_0) := X(\epsilon^{-1}\tau;m_0)$, $\tau\in
[0,\tau_0]$, and $E$ denotes the expectation w.r.t.\ the Brownian
motion $B$ in \eqref{eqsim}.
Let $\xi_\epsilon(\cdot;m_0)$ be as defined in \eqref{eq:6.1}. 
The estimate \eqref{eq:xcx} holds uniformly, namely
\begin{equation}
  \label{eq:xcxu}
\lim_{\epsilon \to 0} \; \sup_{m_0\in \mc N_\eta^{\epsilon,L}}\; 
\bb P \Big(   
\sup_{\tau\in [0,\tau_0]} 
\big| X_\epsilon(\epsilon^{-1} \tau;m_0) - \xi_\epsilon(\tau;m_0) \big| 
> \epsilon^q \Big) = 0.
\end{equation}
Let $\zeta_n := \Xi^{z_0}(\epsilon T_n)$ and denote by 
$\zeta_\epsilon(\cdot;z_0)$ its piecewise linear interpolation as in 
\eqref{eq:6.1}. By \eqref{eq:xcxu} and the continuity of $\Xi^{z_0}$,
\eqref{egu} is proven once we show
\begin{equation}
\label{eguu}
\lim_{\epsilon\to 0} \; \sup_{z_0 \in [-L,L]} \;
\sup_{m_0\in \mc N_\eta^\epsilon(z_0)}\;
\big| \bb E\,F\big(\bam_{\xi_\epsilon(\cdot;m_0)}\big) -
E \,F\big( \bam_{\zeta_\epsilon(\cdot;z_0)}\big)\big|\,=\, 0,
\end{equation}

Given the random variables $\sigma_0,\ldots\sigma_n$, we define the 
sequence $\beta_n$ by the recursive relation $\beta_{n+1} = 
\beta_n+\epsilon T\, b(\beta_n)+\sigma_n$, with $\beta_0 = \xi_0 =
X(m_0)$. The recursive relation \eqref{eq:5.2u}, the bounds
\eqref{t:5.3u} and \eqref{eq:5p1u} imply, by a standard Gronwall argument,
\begin{equation}
\label{eguuu}
\lim_{\epsilon\to 0} \; \sup_{m_0\in \mc N_\eta^{\epsilon,L}}\;
\big| \bb E\,F\big(\bam_{\xi_\epsilon(\cdot;m_0)}\big) -
\bb E\,F\big(\bam_{\beta_\epsilon(\cdot;m_0)}\big)\big|\,=\, 0,
\end{equation}
where $\beta_\epsilon(\cdot;m_0)$ is the piecewise linear interpolation
of the sequence $\beta_n$. 

Recall that $(\Omega,\mc F,\mc F_t, \bb P)$ is the filtered probability
space where the cylindrical Wiener process lives. We denote by 
$(\hat\Omega,\hat{\mc F},\hat{\mc F}_\tau, P)$ the filtered
probability space where the Brownian motion $B$ appearing in 
\eqref{eqsim} lives.
We then set $\widetilde\Omega:= \Omega\times \hat\Omega$, 
$\widetilde {\mc F} := \mc F \times \hat{\mc F}$, 
$\widetilde{\mc F}_t := \mc F_t \times \hat{\mc F}_{\epsilon t}$,
$\widetilde{\bb P} := \bb P \times P$. On this probability space
we define the sequence $\tilde \beta_n$ as
	\begin{equation}
	\label{tY}
\begin{cases}
\tilde\beta_{n+1}=\tilde\beta_n+\epsilon T\, b(\tilde\beta_n)+
 {\displaystyle \sqrt{\frac{s_n}{\epsilon T}}} \:
\big[B(\epsilon T_{n+1})- B(\epsilon T_{n})\big], \\ \tilde\beta_0=\beta_0=\xi_0=X(m_0), \end{cases}
	\end{equation}
where 
$$
s_n \equiv s_n(x_n) := \frac 43\, \bb E\big[ \sigma_n^2 | x_n\big]
= \frac 34 \epsilon \int_0^T\!dt\, \langle \bam'_{x_n},
g^{(x_n)}_{2t}\bam'_{x_n}\rangle. 
$$
Since, conditionally on the centers $x_0,\ldots,x_n$, the random variables $\sigma_0,\ldots,\sigma_n$ are independent Gaussians with variance $\frac 34 s_0,\ldots,\frac 34 s_n$, the sequence $\beta_n$
and $\tilde\beta_n$ have the same law. By \eqref{eguuu}, to prove
\eqref{eguu} it is enough to show that 
\begin{equation}
\label{egue}
\lim_{\epsilon\to 0} \; \sup_{z_0 \in [-L,L]} \;
\sup_{m_0\in \mc N_\eta^\epsilon(z_0)}\;
\tilde{\bb E}\, \big| F\big(\bam_{\tilde\beta_\epsilon(\cdot;m_0)}\big) 
- F\big( \bam_{\zeta_\epsilon(\cdot;z_0)}\big)\big|\,=\, 0,
\end{equation}
where $\tilde \beta_\epsilon(\cdot;m_0)$ is the piecewise linear
interpolation of the sequence $\tilde \beta_n$.  
Set $\varrho_n:=\tilde \beta_n - \zeta_n$; it satifies the recursive
equation
$$
\varrho_{n+1} = \varrho_n + \epsilon T\, \big[b(\tilde\beta_n)] -
b(\zeta_n)  \big] + R^{(1)}_n + R^{(2)}_n, 
$$
where 
\begin{eqnarray*}
R^{(1)}_n & = & \epsilon T \, b(\zeta_n) - \int_{\epsilon T_n}^
{\epsilon T_{n+1}}\! d\tau\, b(\Xi^{z_0}(\tau)), \\
R^{(2)}_n & = & \bigg(\sqrt{\frac{s_n}{\epsilon T}} -1\bigg) 
\big[B(\epsilon T_{n+1})- B(\epsilon T_{n})\big].
\end{eqnarray*} 
Finally, since $\varrho_0 = X(m_0) - z_0$, for each $L>0$ we have
$$
\lim_{\epsilon\to 0} \; \sup_{z_0 \in [-L,L]} \;
\sup_{m_0\in \mc N_\eta^\epsilon(z_0)} 
|\varrho_0| = 0.
$$
By simple estimates on $R^{(i)}_n$, $i=1,2$ and Doob's inequality,
a Gronwall argument shows that, for each $\delta>0$, 
$$
\lim_{\epsilon\to 0} \; \sup_{z_0 \in [-L,L]} \;
\sup_{m_0\in \mc N_\eta^\epsilon(z_0)} \tilde{\bb P}
\Big(\sup_{k\le n_\epsilon(\tau_0)} \; \big|\varrho_k\big| >\delta
\Big) = 0,
$$
which yields \eqref{egue}. 
\qed

\section*{Acknowledgments}

\nopagebreak
It is a great pleasure to thank E.\ Presutti for suggesting us the
problem discussed in this paper and for his collaboration at the initial
stage of the work. We are in debt to L.\ Zambotti for explaining us
Theorem~\ref{1p2}. L.B.\ and P.B.\ acknowledge the partial support of
COFIN-MIUR. S.B. aknowledges the hospitality  at the Mathematics
Department of the University of Rome `La Sapienza'.

\end{document}